\documentclass[twocolumn,times]{aastex62}

\usepackage{natbib}
\usepackage[utf8]{inputenc}
\usepackage[T1]{fontenc}

\newcommand{\foo}[1]{}

\newcommand{\yr}{\>{\rm yr}}
\newcommand{\kms}{\>{\rm km}\,{\rm s}^{-1}}
\newcommand{\masyr}{\>{\rm mas}\,{\rm yr}^{-1}}
\newcommand{\uasyr}{\>\mu{\rm as}\,{\rm yr}^{-1}}

\newcommand{\uas}{\>\mu{\rm as}}

\newcommand{\kpc}{\>{\rm kpc}}

\newcommand{\Msun}{\>{\rm M_{\odot}}}
\newcommand{\degree}{\degr}
\newcommand{\tsim}{\sim\!}

\defcitealias{vdM12a}{vdM12a}
\defcitealias{vdM12b}{vdM12b}
\defcitealias{lindegren18}{L18}
\defcitealias{patel17a}{P17}
\defcitealias{gaiadr2b}{H18}
\defcitealias{salomon16}{S16}

\shorttitle{\textit{Gaia} DR2 Proper Motions of M31 and M33}
\shortauthors{}

\begin{document}

\title{\textbf{\Large First \textit{Gaia} Dynamics of the Andromeda System:\\ DR2 Proper Motions, Orbits, and Rotation of M31 and M33}}

\author[0000-0001-7827-7825]{Roeland P. van der Marel}
\affiliation{Space Telescope Science Institute, 
             3700 San Martin Drive, 
             Baltimore, MD 21218, USA}
\affiliation{Center for Astrophysical Sciences, Department of Physics \& Astronomy, 
             Johns Hopkins University, 
             Baltimore, MD 21218, USA}
             
\author[0000-0003-4207-3788]{Mark A. Fardal}
\affiliation{Space Telescope Science Institute, 
             3700 San Martin Drive, 
             Baltimore, MD 21218, USA}

\author[0000-0001-8368-0221]{Sangmo Tony Sohn} 
\affiliation{Space Telescope Science Institute, 
             3700 San Martin Drive, 
             Baltimore, MD 21218, USA}

\author[0000-0002-9820-1219]{Ekta Patel}
\affiliation{Department of Astronomy, University of Arizona,
             933 North Cherry Avenue, 
             Tucson, AZ 85721, USA}

\author{Gurtina Besla}
\affiliation{Department of Astronomy, University of Arizona,
             933 North Cherry Avenue, 
             Tucson, AZ 85721, USA}

\author[0000-0003-4922-5131]{Andr\'{e}s del Pino}
\affiliation{Space Telescope Science Institute, 
             3700 San Martin Drive, 
             Baltimore, MD 21218, USA}
             
\author[0000-0001-9525-3673]{Johannes Sahlmann}
\affiliation{Space Telescope Science Institute, 
             3700 San Martin Drive, 
             Baltimore, MD 21218, USA}

\author[0000-0002-1343-134X]{Laura L. Watkins}
\affiliation{Space Telescope Science Institute, 
             3700 San Martin Drive, 
             Baltimore, MD 21218, USA}

\begin{abstract}
The 3D velocities of M31 and M33 are important for understanding the evolution and cosmological context of the Local Group. Their most massive stars are detected by \textit{Gaia}, and we use Data Release 2 (DR2) to determine the galaxy proper motions (PMs). We select galaxy members based on, e.g., parallax, PM, color-magnitude-diagram location, and local stellar density. The PM rotation of both galaxies is confidently detected, consistent with the known line-of-sight rotation curves: $V_{\rm rot} = -206 \pm 86 \kms$ (counter-clockwise) for M31, and $V_{\rm rot} = 80 \pm 52 \kms$ (clockwise) for M33. We measure the center-of-mass PM of each galaxy relative to surrounding background quasars in DR2. This yields that $(\mu_{\alpha*},\mu_{\delta})$ equals $(65 \pm 18 , -57 \pm 15) \uasyr$ for M31, and $(31 \pm 19 , -29 \pm 16) \uasyr$ for M33. In addition to the listed random errors, each component has an additional residual systematic error of $16 \uasyr$. These results are consistent at $0.8\sigma$ and $1.0\sigma$ with the (2 and 3 times higher-accuracy) measurements already available from \textit{Hubble Space Telescope} (\textit{HST}) optical imaging and VLBA water maser observations, respectively. This lends confidence that all these measurements are robust. The new results imply that the M31 orbit towards the Milky Way is somewhat less radial than previously inferred, $V_{\rm tan, DR2+HST} = 57^{+35}_{-31} \kms$, and strengthen arguments that M33 may be on its first infall into M31. The results highlight the future potential of \textit{Gaia} for PM studies beyond the Milky Way satellite system.
\end{abstract}

\keywords{galaxies:kinematics and dynamics --- Local Group -- proper motions}

\section{Introduction}

The Milky Way (MW), Andromeda (M31) and Triangulum (M33) galaxies are the three most massive members of the small group of galaxies commonly referred to as the Local Group (LG). Together these spiral galaxies make up the majority of the mass in the LG, which has been estimated to weigh approximately 3--$5 \times 10^{12} \Msun$ \citep[e.g.,][hereafter \citetalias{vdM12a}]{gonzalez14, vdM12a}. 

As our nearest laboratory for testing theories of galaxy formation and evolution, the LG and its constituents are the best examples of hierarchical structure formation and large-scale structure. Studies of galactic archeology and near-field cosmology have made tremendous progress in recent decades, and this has placed the LG in a proper cosmological context. However, much of this work was carried out without detailed knowledge of the three-dimensional (3D) velocity vectors of LG objects. At the distances of these objects, the proper motions (PMs) are small and hard to measure with traditional techniques. As a result, the relative motion of M31 with respect to the MW has been a matter of debate. This motion is central to our understanding of both the assembly and current state of the LG \citep[e.g.,][]{foreroromero13, peebles13} and its future evolution \citep[][hereafter \citetalias{vdM12b}]{cox08, vdM12b}. 

PM measurements are also an essential ingredient for a better understanding of the dynamics of satellite galaxies and tidal streams. This has been successfully explored in the halo of the MW system \citep[e.g.,][]{paw13,sohn15}, but measurements for the halo of the Andromeda system are still lacking. Also, PM measurements can reveal the internal rotation and structure of galaxies. Reports of this date back a century with the (discredited) work of van Maanen \citep[reviewed in][]{hether72}. This has now become possible though, but to date the technique has only been realized for the Large (LMC) and Small (SMC) Magellanic Clouds \citep{vdmnk14,vdm16,nie18,ziv18}. Among other things, this makes it possible to obtain kinematic distance estimates when combined with LOS velocity data \citep{oll00}. 

The line-of-sight (LOS) velocity of M31 was first determined by Slipher using observations performed in 1912 \citep{slipher13}. Exactly one century later, observations with the \textit{Hubble Space Telescope} (\textit{HST}) were used to report for the first time the absolute PM \citep[][\citetalias{vdM12a}]{sohn12}. \textit{HST} observed three fields of stars in M31 over a 5--7 year baseline to obtain a measurement with an accuracy per coordinate of 12 $\mu$as yr$^{-1}$ ($\sim 45 \kms$). 

Alternatively, the transverse velocity $V_{\rm tan}$ of M31 can be estimated by indirect dynamical methods based on modeling the LOS velocities of M31 or LG satellites. A collection of methods was presented in \citet{vdMG08}, and their implications were subsequently refined with more recent data in \citetalias{vdM12a}. These methods assume little more than non-rotating equilibrium. \citet[][hereafter \citetalias{salomon16}]{salomon16} used a variation on one of these methods and applied it to a larger sample of satellite galaxies with more precise distance measurements. Their method makes more specific assumptions about the dynamical equilibrium of the satellites, but was verified using cosmological simulations. All methods yield a fairly consistent $V_{\rm tan}$, with a method-dependent uncertainty of $\sim 60$--$90 \kms$ per coordinate. 

The PM measured with \textit{HST} differs from the $V_{\rm tan}$ implied by the indirect dynamical methods. In case of the \citetalias{salomon16} values, the difference is 130--$140 \kms$ in each coordinate, with an uncertainty of $\sim 80 \kms$. This is significant at the $1.9\sigma$ level. \citetalias{vdM12a} posited that different methods probably have different systematics, so that the most accurate estimate is obtained by averaging the direct PM measurement with the indirect dynamical results. Either with or without this averaging, the resulting velocity is statistically consistent with a direct radial (head-on collision) orbit for M31 towards the MW, implying a future collision and merging of the two galaxies \citepalias{vdM12b}. By contrast, \citetalias{salomon16} adopted their indirect dynamical estimate as the preferred one, and hence argued that $V_{\rm tan}$ is in fact $165 \pm 62 \kms$, in which case the LG may not be a bound system. These discrepancies clearly impact our understanding of the dynamics of the LG. 

The situation is different for M33. The PM of M33 was determined using VLBA water maser observations by \citet{brunthaler05}. VLBA has very high intrinsic spatial resolution, unlike \textit{HST}, which has to measure PMs at levels below one-hundredth of a pixel. The VLBA determination is therefore likely to be robust. However, the motion of M33 relative to M31 is less well-constrained, due to the uncertainties in the PM of M31.

The M33-M31 orbit is interesting for multiple reasons. Observations of M33 have provided evidence for warps in its outer stellar and gaseous disks \citep{rogstad76, corbelli97, putman09, corbelli14, kam17}. Tidal streams have been detected as well \citep{mcconnachie09}. By aiming to match these morphological features in M33 via simulations, it is possible to constrain the allowed M33 orbits and M31 PM values \citep{loeb05, vdMG08}. \citet{mcconnachie09} find that the stellar debris around M33 can be formed through a recent (< 3 Gyr ago), close (< 55 kpc) tidal interaction with M31. \citet{semczuk18} argue that the \citetalias{salomon16} estimate of M31's $V_{\rm tan}$ is more consistent with this scenario than the \textit{HST} PM measurement, but they did not explore the full space of orbits allowed within the uncertainties.

The M31 \textit{HST} and M33 VLBA PM measurements can be combined to determine both the future orbital evolution \citepalias{vdM12b} and past orbital history of the M33-M31 system. \citet[][hereafter \citetalias{patel17a}]{patel17a} calculated the plausible orbital histories for M33 to determine which orbital solutions are allowed within the observational uncertainties. They concluded that M33 is either on its first infall into the halo of M31 or that it is on a long-period orbit ($\sim$ 6 Gyr) where it completed a pericentric approach at a distance of $\sim$100 kpc. First infall orbits are in fact cosmologically expected for satellites in this mass range at the present epoch \citep[][\citetalias{patel17a}]{bk11}.

New observational evidence for the PMs of M31 and M33 is highly desirable to discriminate between the various scenarios discussed above. The Data Release 2 (DR2) \citep{gaiadr2a} of the \textit{Gaia} mission \citep{gaiaMission} provides an opportunity for progress. The \textit{Gaia} mission is optimized for studies of the MW \citep{ktz18} and its satellite system \citep[][hereafter \citetalias{gaiadr2b}]{simon18,fritz18,kallivayalil18,mas18,gaiadr2b}. However, rare supergiant stars in star-forming regions can be bright enough to be detected by \textit{Gaia} even at the distance of the Andromeda system. We therefore present here the first \textit{Gaia} study of the dynamics of the Andromeda system, focusing on the PMs of M31 and M33 as revealed by the DR2.

\begin{figure*}
\gridline{
\fig{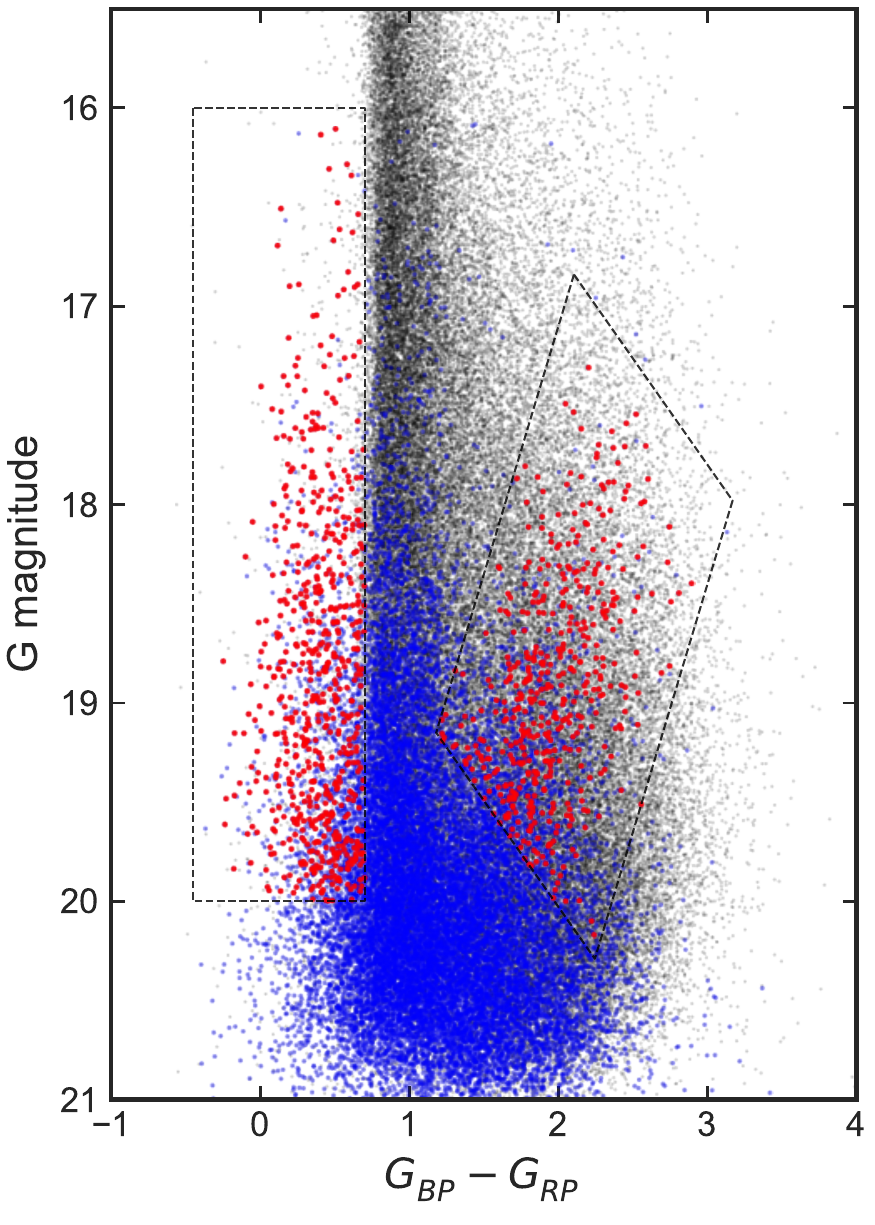}{0.24\textwidth}{(a)}
\fig{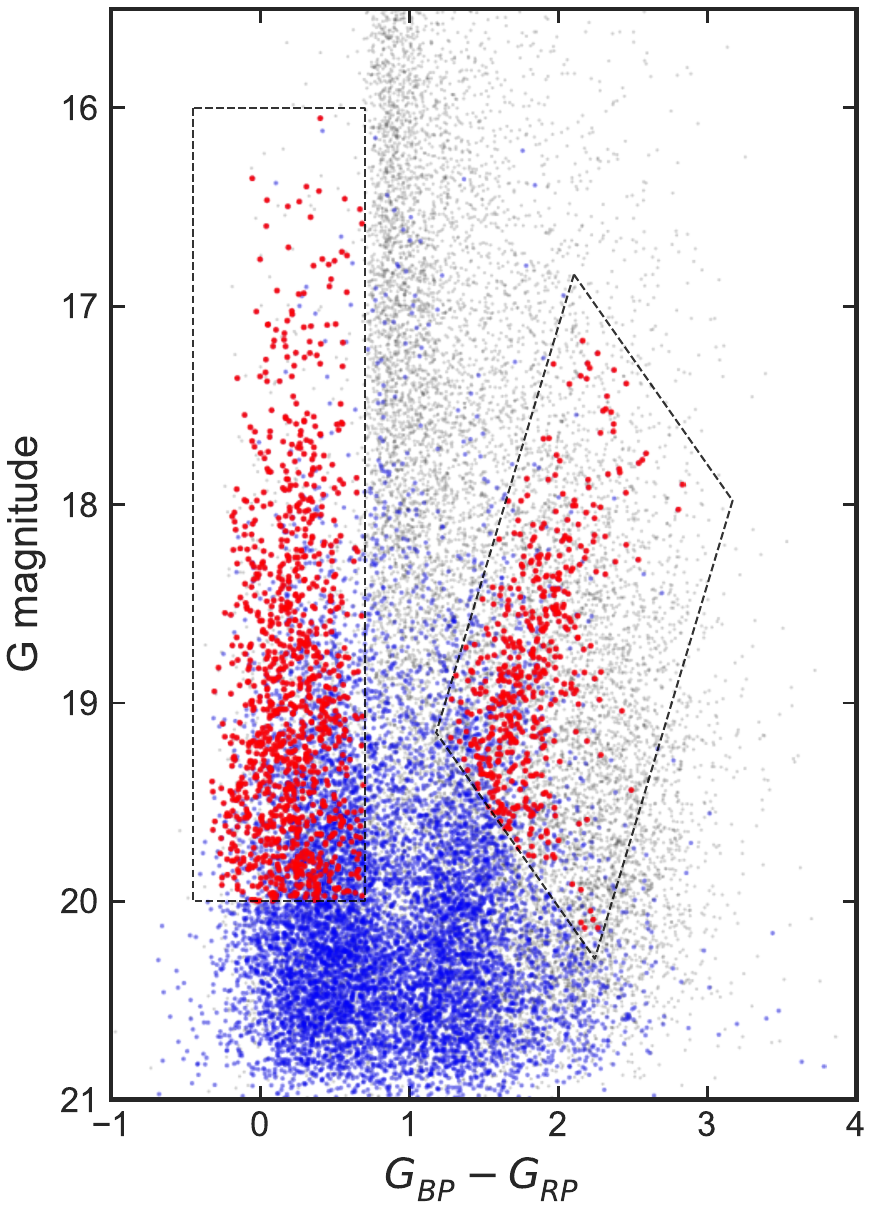}{0.24\textwidth}{(b)}
\fig{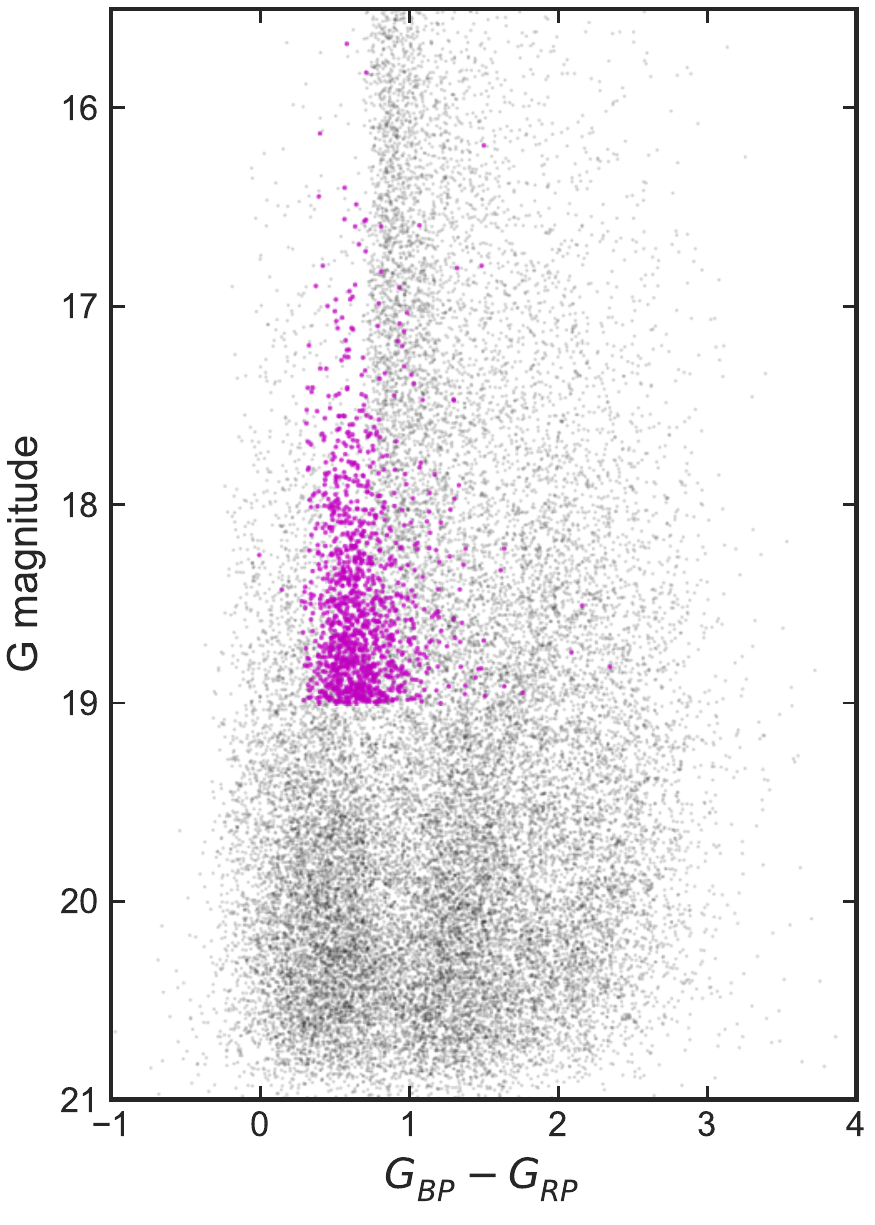}{0.24\textwidth}{(c)}
\fig{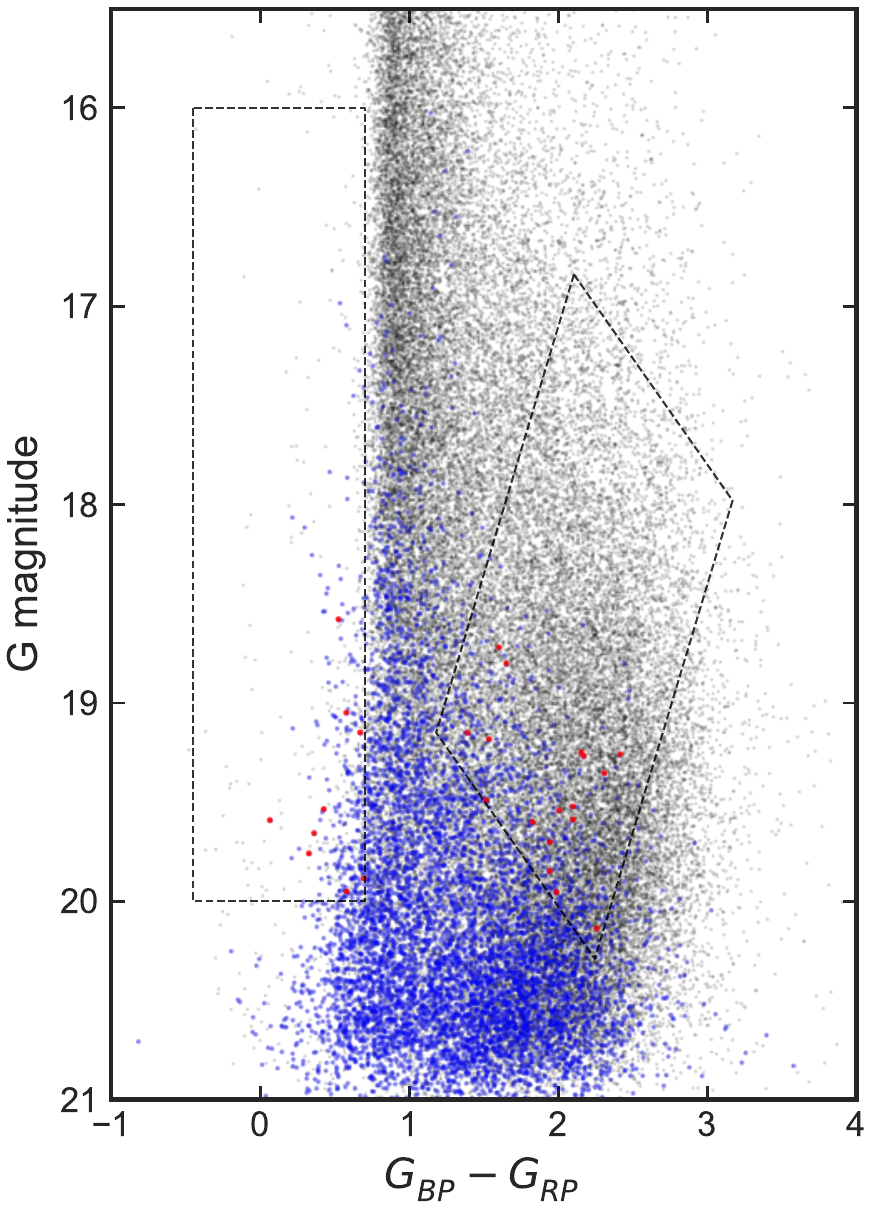}{0.24\textwidth}{(d)}}
\caption{CMD of (a) M31; (b) and (c) M33; and (d) four background comparison regions of M33 combined. Gray points show all \textit{Gaia} DR2 sources within the circular extraction region with valid proper motions. Blue points in panels (a), (b) and (d) show sources also passing the parallax and loose proper motion cuts discussed in the text. Red points in panels (a), (b) and (d) show sources that also pass the cuts on astrometric fit quality, photometry, elliptical galaxy boundary, local spatial density, and CMD position for membership in the final sample; the selection boxes used for the CMD cuts are shown in the panels. Magenta triangles in panel (c) show the quasar sample for M33, selected as described in Section~\ref{sec:qsosample}, used to correct the M33 astrometric reference frame as described in Section~\ref{sec:method} and Appendix~A.\label{fig:cmd}}
\end{figure*}

The available accuracies with DR2 are not yet competitive with either \textit{HST} or VLBA, but they are close. So by themselves, they cannot yet resolve most of the aforementioned questions. However, they have the potential to discriminate some opposing models and scenarios, and they provide an independent consistency check.  For example, both the M31 measurement with \textit{HST} and the M33 measurement with VLBA use small areas within these galaxies, and must correct for the internal kinematics within these galaxies which is a potential source of systematic error.  \textit{Gaia} observes the entire disk of each galaxy and thus is more robust in this respect.  \textit{Gaia} can can also help check for purely instrumental biases in the other measurements.
Moreover, it is possible to measure the PM rotation of both galaxy disks. The present study derives the current constraints from \textit{Gaia} in these areas, and highlights the potential for further progress with future \textit{Gaia} data releases.

The outline of this paper is as follows. Section 2 discusses the selection of \textit{Gaia} DR2 stars in the target galaxies, and the selection of background quasars used for (partial) correction of systematic PM uncertainties in the \textit{Gaia} DR2 catalog. Section 3 analyzes the samples to determine the disk rotation and center-of-mass (COM) PM of each galaxy, and the implied Galactocentric velocities. The results are compared to previous measurements and estimates in the literature. Section 4 discusses the implications for our understanding of the LG. Section 5 summarizes the results. Appendices discuss the systematic uncertainties in our measurements, and the types of stars detected by \textit{Gaia} in M31 and M33.

\section{\textit{Gaia} DR2 Data Samples}

\subsection{M31 and M33 Sample Selection}
\label{sec:sample}

\begin{figure*}
\gridline{
\fig{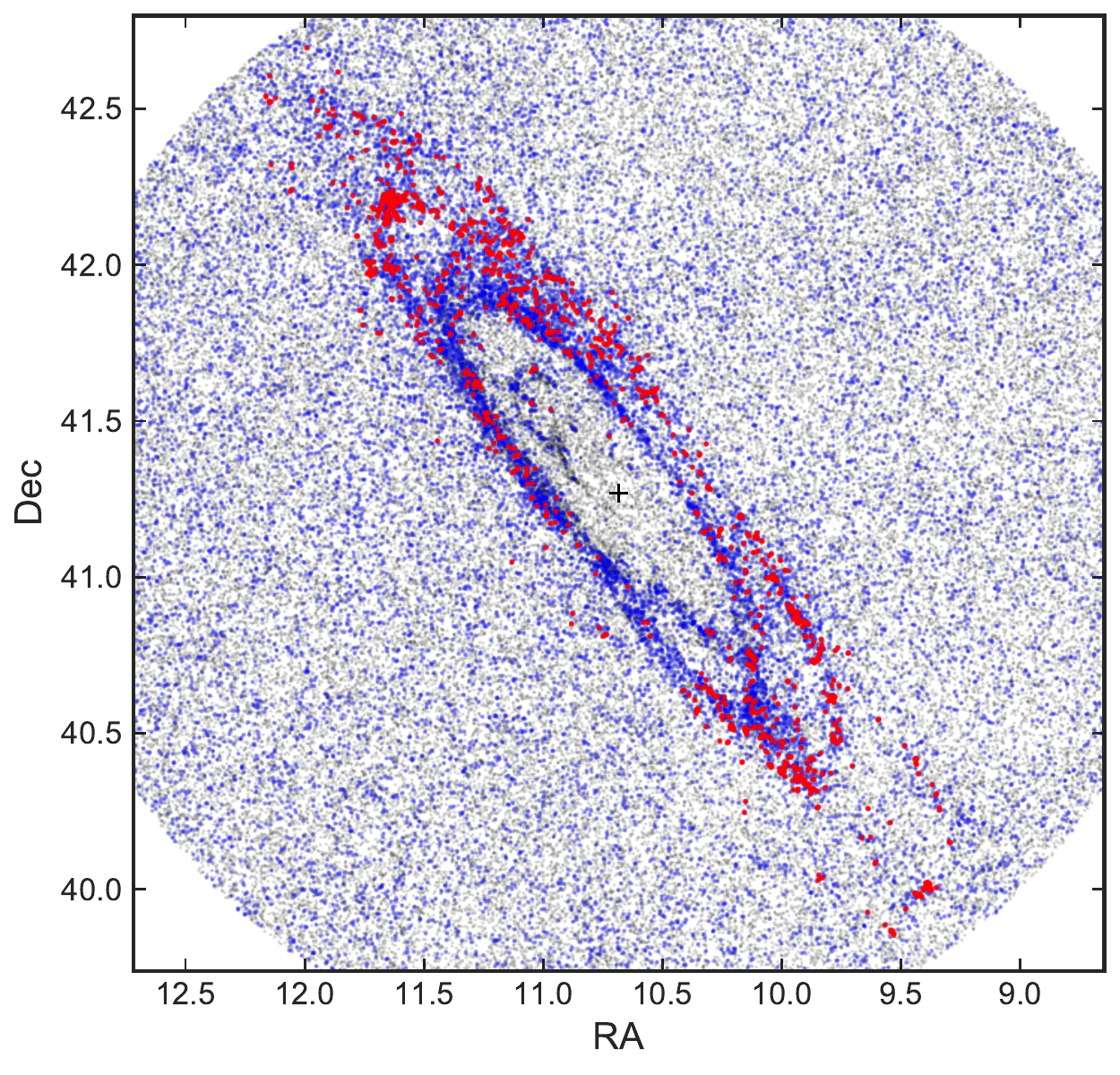}{0.5\textwidth}{(a)}
\fig{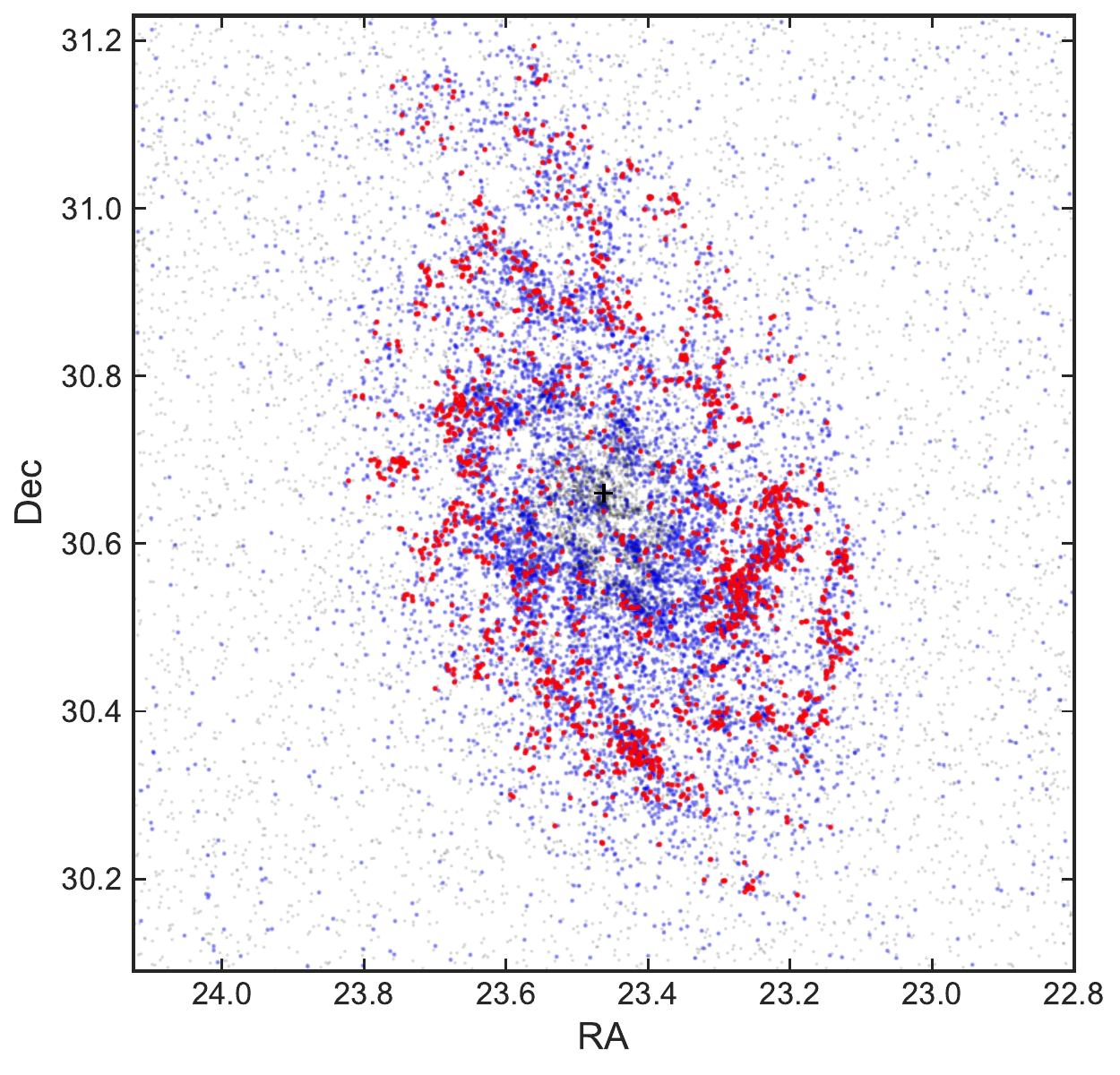}{0.5\textwidth}{(b)}}
\caption{Spatial distribution of the \textit{Gaia} DR2 sources identified in Figure~\ref{fig:cmd}, using the same color-coding, for: (a) M31; and (b) M33. The left panel subtends a linear size that is $\sim 3$ times larger than the right panel.\label{fig:spatial}}
\end{figure*}

The actively star-forming regions of both M31 and M33 produce a large number of bright young stars along with nebular emission. Both galaxies were easily visible in sky maps of the \textit{Gaia} DR1 catalog, and displayed the characteristic spatial pattern of star-forming regions in the individual galaxies, such as a strong concentration in M31's 10~kpc star-forming ring.  Plotting the sources on SDSS images showed the vast majority of sources were point sources rather than patches of nebular emission. A cross-match of DR1 sources with the LGGS source catalog \citep{LGGS16} confirmed that the color-magnitude diagram (CMD) is consistent with that expected for supergiants at the distance of the Andromeda system. Based on this pre-release assessment, we extracted \textit{Gaia} DR2 sources from a circular region around each galaxy, of radius $1.8 \degree$ for M31 and $1.0 \degree$ for M33. We removed all sources with missing proper motions. Figure~\ref{fig:cmd} shows in grey the CMDs of the remaining sources, and  Figure~\ref{fig:spatial} illustrates their spatial distributions.

We proceeded to impose various sample cuts intended to screen out contaminants and bad measurements. We removed sources with parallax values inconsistent with the distance of the Andromeda system ($\tsim 800 \kpc$) at greater than the $2\sigma$ level, using the global parallax zero-point $\varpi_0 = -0.03 \masyr$ estimated by \citet[][hereafter \citetalias{lindegren18}]{lindegren18}. We also removed sources outside of an initial color-magnitude box defined by 
$-1.0 < G_{BP}-G_{RP} < 4.0$ and $G > 16$.
Moreover, we imposed very loose proper motion requirements of
$|\mu_{\alpha*}| < 0.2 \masyr + 2.0 \sigma_{\mu_{\alpha*}}$ and
$|\mu_{\delta}|  < 0.2 \masyr + 2.0 \sigma_{\mu_{\delta}}$. At the distance of the Andromeda system, this removes sources with velocities that differ by $\gtrsim 500 \kms$ from those of M31 and M33. These choices screen out most foreground sources. The remaining sources are shown in blue in Figures~\ref{fig:cmd} (panels a, b, and d) and~\ref{fig:spatial}. 

We then removed sources with bad astrometric fits (a few percent of the overall catalog) following equation~C.1 in \citetalias{lindegren18}: defining $u \equiv (\mbox{\tt astrometric\_chi2\_al} /$ $\mbox{\tt astrometric\_n\_good\_obs\_al} - 5)^{1/2}$, we require $u < 1.2 \times \max(1, \exp(-0.2(G-19.5)))$. Moreover, we selected only those sources that fall within an ellipse on the sky outlining the star-forming regions of each galaxy. The major axes were chosen as $1.8$ deg for M31 and $0.6$ deg for M33, with the shapes and orientations of the ellipses consistent with the known viewing angles of the galaxy disks (see Section 3.1). 

The CMDs of the remaining sources in each galaxy exhibit two plumes forming a V pattern, corresponding predominantly to blue and red supergiant stars, and some blue main-sequence stars (for more detail on the nature of these sources we refer the reader to Appendix~C). This pattern becomes less distinct towards the center of each galaxy, particularly in M31.  The flux excess factor $E \equiv \mbox{\tt phot\_bp\_rp\_excess\_factor}$, which compares the $G$ magnitude to the value expected from the $G_{BP}$ and $G_{RP}$ magnitudes, also takes on increasingly high values towards the center for most sources.  \citetalias{lindegren18} give a cut on this quantity in equation C.2 that improves the behavior in the CMD. If applied in our case, this would leave very few target stars.  We are not sure of the exact cause for the rising flux excess values toward the center of these galaxies, but it may have to do with either scattered light in the BP/RP optical path or poor estimation of the background levels in the BP/RP photometers (as mentioned by \citealp{are18}).  It seems likely that it could affect the measured colors without substantial effect on the astrometry.  We therefore use a similar but more tolerant cut of
$1 + 0.015 (G_{BP} - G_{RP})^2 < E < 1.5[1.3 + 0.06 (G_{BP} - G_{RP})^2]$,
intended to limit sources to those with reliable enough $G_{BP}$ and $G_{RP}$ photometry to leave selection via our broad CMD cuts relatively unaffected.  This preferentially suppresses sources in the central regions of each galaxy.

\begin{figure*}[tp]
\gridline{
\fig{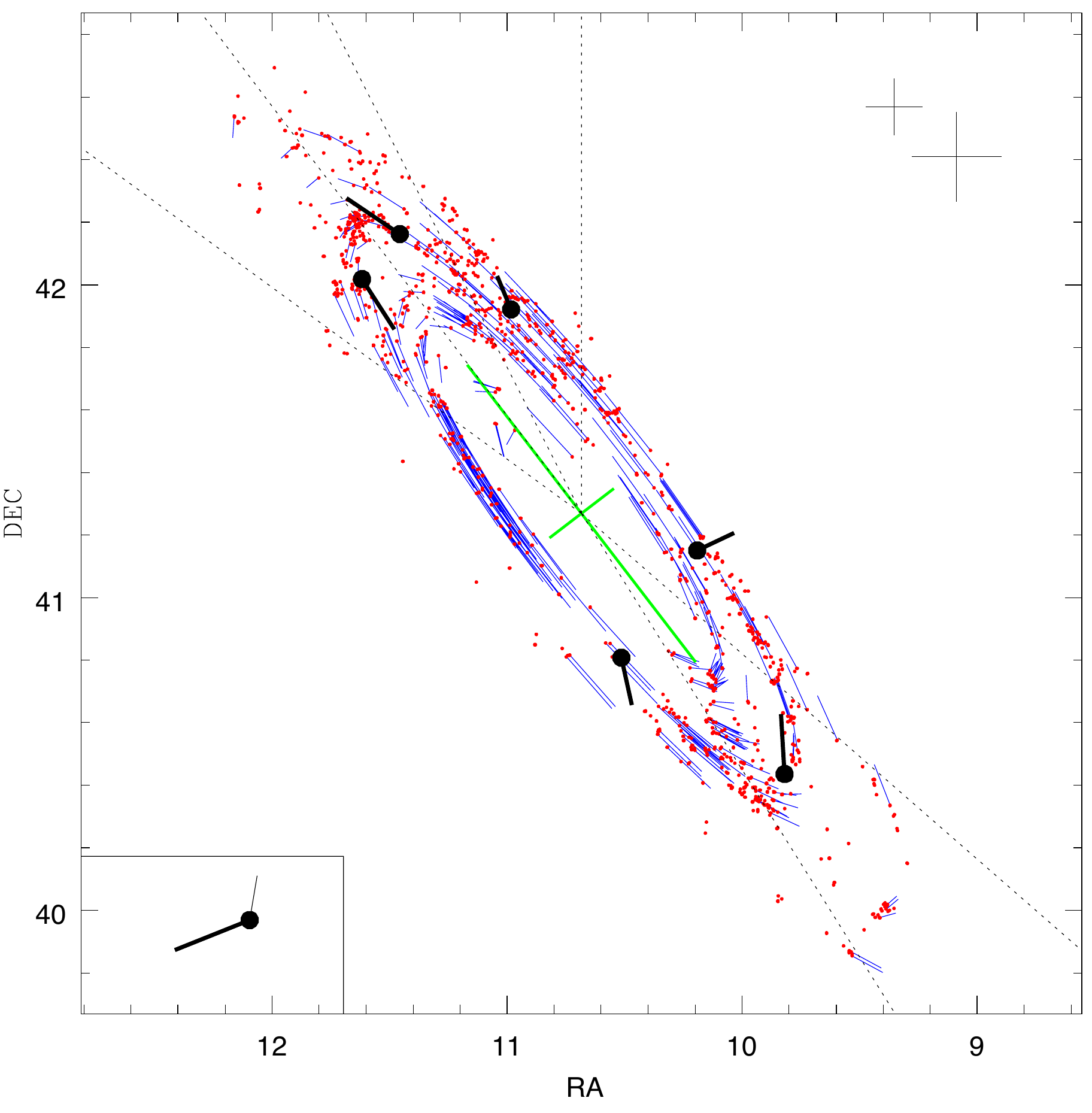}{0.485\textwidth}{(a)}
\fig{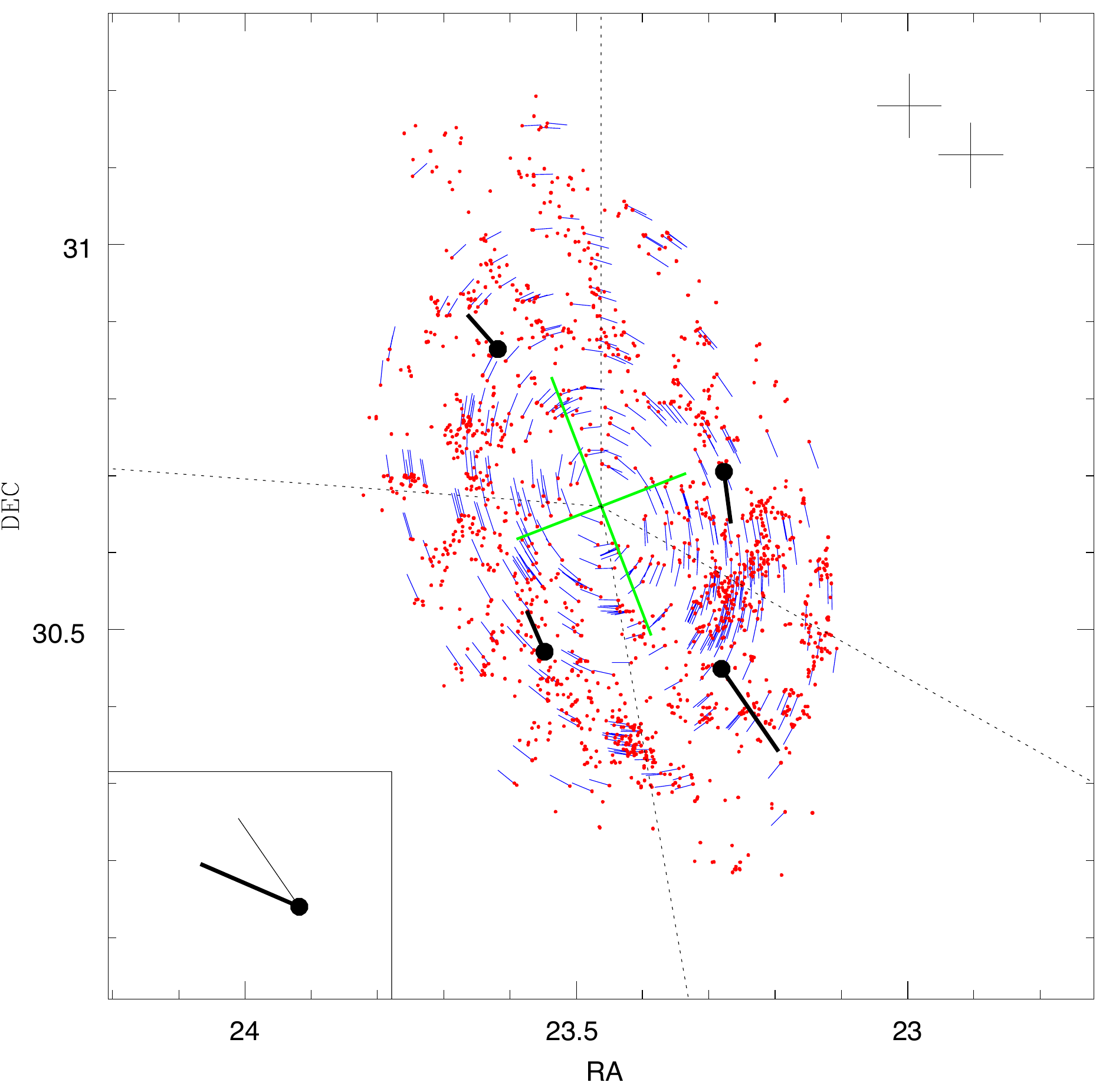}{0.5\textwidth}{(b)}}
\caption{\label{fig:PMrotation}Proper Motion kinematics of M31 (left) and M33 (right). Red points are the target sources selected from the \textit{Gaia} DR2 (same as in Figure~\ref{fig:spatial}). Blue line segments show the PMs {\it predicted} by the best-fit rotating disk models, determined as described in Section~3, for sources brighter than G=18.5. Black line segments show the weighted averages of all the {\it observed} PMs, obtained by binning the sources in 6 (M31) or 4 (M33) sectors in position angle, indicated with thin dotted lines, with equal numbers of stars per sector. The rotation of each galaxy is visually apparent. The best-fit COM PMs were subtracted from each of the displayed PM vectors. These COM PMs are shown as thick line segments in the insets on the bottom left. The average PM of surrounding quasars is shown as thin line segments. The difference between these vectors corresponds to our final corrected COM PM estimates. In each case, the PM direction starts at the dot and moves along the line segment. The error bars in the top right show respectively, from left to right: the final uncertainty on the corrected COM PM determination; and the median PM uncertainty for the sector averages. The median PM uncertainty for the individual sources in the sample is about 25 times larger than the former. The green line segments show the adopted position angles and projected ratio of the major and minor axes.}
\end{figure*}

Between the blue and red plumes in the CMD, there is a vertical plume of stars indicating contamination by foreground main-sequence turnoff stars. We avoid this region in our final sample selection, by allowing only sources that fall in one of two disjoint selection regions (shown in Figure~\ref{fig:cmd}), namely blue sources with
$-0.4 < G_{BP} - G_{RP} < 0.70$ and
$16 < G < 20$,
and red sources with 
$22.1 < G + 2.50 (G_{BP} - G_{RP}) < 25.9$ and
$14.586 < G - 1.071 (G_{BP} - G_{RP}) < 17.886$.

Any remaining contaminants should not be spatially correlated with the high-density star-forming regions in the target galaxies. To further reduce contamination, we therefore created a kernel density estimate. For this estimate we used all sources irrespective of CMD position, which reduces noise (tests show that M31 or M33 sources dominate every heavily-populated part of the CMD, while remaining contaminants should be smoothly distributed on the sky). We used individual smoothing lengths for each galaxy chosen to pick out the dominant scale of star-forming regions. We then kept for the final astrometric analysis only those sources passing a density threshold set individually for each galaxy. The effects of the full sequence of sample cuts are illustrated in the CMD plots in Appendix~C.

The final samples thus selected contain 1084 sources for M31, and 1518 sources for M33. These sources are shown in red in Figures~\ref{fig:cmd} (panels a, b, and d) and~\ref{fig:spatial}. The two disjoint groupings in the CMDs reflect the choice of selection regions. The spatial distributions of the sources clearly reflects the morphology of the star-forming regions. In fact, this morphology is also evident in the distribution of sources shown in blue that did {\it not} pass all of the cuts. This implies that there are bona-fide members of the target galaxies that were excluded from the sample, so as to guarantee a minimum amount of non-member contamination.   

The few brightest stars at $G \sim 16$ have PM uncertainties approaching $\sim 100 \uasyr$. However, the PM uncertainty at the median sample brightness of $G \sim 19$ is $\sim 600 \uasyr$. Here and henceforth we assume a distance $D = 770 \kpc$ for M31 and $D = 794 \kpc$ for M33 (\citetalias{vdM12a, vdM12b} and references therein). At these distances, $1 \uasyr$ corresponds to $3.65$ and $3.76 \kms$, respectively. Therefore, the {\it individual} PM uncertainties are too large to assess the galaxy's kinematics. However, by averaging or model fitting, the uncertainties can be reduced to interesting levels. Figure~\ref{fig:PMrotation} shows with black line segments the weighted average PMs obtained by binning the sources in 4--6 sectors in position angle, with equal numbers of stars per sector\footnote{The azimuthal variation in the PM averages depends somewhat on the specific binning adopted, which is fairly arbitrary. However, this binning is only used for visual illustration, and not for the quantitative analysis of Section~\ref{sec:method}.}. The COM PM estimates from Section 3 (which are similar to the weighted average PMs of the entire sample) were subtracted. The rotation of each galaxy is now visually apparent.   

To assess the impact of residual contamination on the PM analysis, we extracted from the DR2 another 10 additional regions for each galaxy. For M31, these regions were centered at $l = 121.2\degree + 3.5\degree \, k$ and and $b=-21.6\degree$, with $k = -7 \ldots -3, 3 \ldots 7$. For M33, they were centered at $l=133.6\degree + 3.5\degree \, k$ and and $b=-31.3\degree$, with the same values of $k$. We then re-centered these sources on the target galaxies by converting their positions to tangent-plane coordinates around the center of the background region, and then converting back to sky coordinates using the target galaxy as the center. When testing the effect of contamination, the same cuts (including spatial cuts) were applied as for the actual target galaxy data. The right panel of Figure~\ref{fig:cmd} shows the CMD for four of the combined M33 comparison regions. The paucity of red points in this CMD implies that our final blue and red CMD selection regions are relatively uncontaminated, particularly the blue region. Using a weighted PM average, we estimate  that inclusion of the remaining contamination adds a bias of absolute value $\lesssim 2 \uasyr$ to the measured PM of each galaxy, with a scatter of a comparable amount. This bias is a factor 5--7 below the random PM errors obtained by averaging over all sources, and can therefore be safely ignored. 

\subsection{Quasar Sample Selection}
\label{sec:qsosample}

The zero-point of the Gaia DR2 proper motion reference frame clearly
varies over the sky, as described in \citet{lindegren18}, \citet{gaiadr2b}, and \citet{mignard18}. To assess and (partially) correct for PM systematics in the DR2, as described in Section 3.1 and Appendix A, we also require the measured PMs for samples of quasars. First, we extracted \textit{Gaia} DR2 sources within $10$ deg radius around the target galaxies that matched the criteria outlined in Equations~13(i)--(iv) of \citetalias{lindegren18}: (1) \texttt{astrometric\_matched\_observations} $\geq 8$; (2) $\sigma_{\varpi} < 1$~mas; (3) $|\varpi / \sigma_{\varpi}| < 5$; (4) $(\mu_{\alpha*}/\sigma_{\mu_{\alpha*}})^{2} + (\mu_{\delta}/\sigma_{\mu_{\delta}})^{2} < 25$, where $\sigma$ denotes the errors of each parameter as provided in the \textit{Gaia} DR2 catalog. This effectively filters out many Galactic stars and reduces their contamination rate. Subsequently, the remaining sources were cross matched with the AllWISE AGN/QSO catalog of \citet{sec15,sec16} within a matching  radius of 1~arcsec. Areas close to the disks of M31 and M33 were excluded to avoid contamination.\footnote{We note that the AllWISE AGN/QSO catalog contains many stars that belong to the M31 and M33 disks as revealed by the concentration of sources near these galaxies.} After visually inspecting the spatial distribution of both the \textit{Gaia} DR2 and AllWISE sources, we retained only radii in excess of $3.0$ deg and $1.5$ deg from the centers of M31 and M33, respectively. Since the risk for contamination may be highest at the faintest magnitudes, where the PM measurement uncertainties are large anyway, we only retained sources brighter than $G = 19$. In calculating weighted PM averages for the samples we iteratively rejected sources with PMs that are discrepant at the $>3\sigma$ level. We also experimented with sample restrictions using CMD selections, but found that this did not significantly affect the average PMs. The total numbers of quasars used for M31 and M33 are 866 and 1,174, respectively.

Figure~\ref{fig:cmd}c shows in magenta the CMD of the quasar sample around M33; the quasar sample around M31 occupies the same CMD region. The quasar sample has mean $G=18.4$, $BP-RP=0.7$. The selected M33 stars have mean $G=18.9$, $BP-RP=0.7$, while the selected M31 stars have mean $G=18.9$, $BP-RP=1.1$. Therefore, the quasars we use as a reference frame are fairly well-matched in magnitude and color to the stellar targets. This is important, because it is not well known to what extent the \textit{Gaia} DR2 proper motion zero-point depends on source color or magnitude. The running averages of the mean quasar {\it parallax} versus magnitude or color in \citet{lindegren18} and \citet{chen18} show small variations not much in excess of the statistical noise. From the small offsets in magnitude or color between our target and quasar samples, potential offsets in the parallax zero-point are $\ll 10 \uas$. In DR2, the statistical or systematic proper motion errors are typically $1.7 \yr^{-1}$ times the corresponding parallax errors (see equations~16 and 17 and table~B1 of \citealp{lindegren18}). We thus expect systematic proper errors of $\ll 15 \uasyr$ from color or magnitude dependencies. This is below the known systematic and statistical errors we quantify in our analysis below, and we therefore neglect these errors.

\section{Proper Motion Analysis and Results}

\subsection{Analysis Methodology}
\label{sec:method}

We use the same methodology and equations to fit the M31 and M33 PM fields as was used for the case of the LMC in \citet{vdmnk14} and \citet{vdm16}. The stars are assumed to reside in a flat disk, moving on circular orbits around the COM. The disk orientation is governed by the inclination angle $i$ and the position angle $\Theta$ of the line of nodes, and is assumed to be fixed with time (i.e., no precession or nutation of the spin axis). The influence of viewing perspective (i.e., the lines of sight toward different points in the disk not being parallel, and the distances not being equal) is taken into account through full spherical trigonometry. The \textit{Gaia} DR2 data are not of sufficient quality to meaningfully constrain the position of the COM, the viewing angles, the distance $D$, or the LOS velocity $V_{\rm LOS}$ of the COM.\footnote{The latter is in principle constrained by the PM field, because the approaching velocity of these galaxies causes them to grow in size on the sky, implying an outward radial component in the PM field. By quantifying this and comparing it to the observed COM LOS velocity one can obtain a kinematic distance estimate \citep[see][]{vdmnk14}.} So we keep these quantities fixed to the values implied by existing photometric and LOS velocity studies. For M31 we use COM position (RA,Dec) $ = (10.68333, 41.26917)$ deg, $v_{\rm LOS} = -301 \kms$, $i = 77.5$ deg, and $\Theta = 37.5$ deg; for M33 we use (RA,Dec) $ = (23.46250, 30.6602)$ deg, $v_{\rm LOS} = -180 \kms$, $i = 49.0$ deg, and $\Theta = 21.1$ deg (\citetalias{vdM12a, vdM12b}, and references therein). The \textit{Gaia} DR2 data are also not of sufficient quality to meaningfully constrain the shape of the rotation curve, so we assume the rotation curve to be flat.

With this model, the PM field is determined by only three free parameters, the PM $(\mu_{\alpha*},\mu_{\delta})$ of the COM, and the constant rotation velocity $V_{\rm rot}$ in the disk. We vary the model parameters to minimize the $\chi^2$ of the fit to the data, taking into account the PM correlations \mbox{\tt pmra\_pmdec\_corr} of individual stars given in the DR2 (which have a median value of $\sim 0.25$), so as to obtain the best-fitting values. We increase the measurement uncertainties of the sample stars by a factor of $1.10$, since DR2 verification papers have found that the uncertainties may be underestimated by this factor \citep[\citetalias{lindegren18},][]{are18}. We iterate with outlier rejection, to remove the 4\% of data points that yield the highest residuals. The final minimum $\chi^2$ is slightly below the number of degrees of freedom, as expected given the outlier rejection. We then create pseudo-data sets in Monte-Carlo fashion. These have the same stars as the target samples, but now with PMs set by adding deviates drawn at random from the measurement uncertainties to the model predictions. These pseudo-data are analyzed like the real data to determine the uncertainties on the inferred model parameters.  We found that using bootstrap sampling instead of Monte
Carlo did not significantly change the results, which indicates the adopted statistical errors are robust.

\citetalias{lindegren18} and \citetalias{gaiadr2b} reported that the DR2 catalog has spatially correlated PM errors. These systematic errors have two distinct components, a small-scale and a large-scale component. The small-scale component appears as a pattern noise related to the \textit{Gaia} scan pattern, with a characteristic scale of $\sim 1$ deg. \citetalias{gaiadr2b} show this pattern, e.g., in their figures 16 and A10, and discuss it in their section 4.1. They quote an RMS amplitude of $35 \uasyr$, while \citetalias{lindegren18} estimate a somewhat larger amplitude. However, the sample stars for both M31 and M33 extend over a significant region (see Figure~\ref{fig:spatial}). Hence, the small-scale PM variations are expected to largely average out in our model fitting. In Appendix~B we estimate the possible remaining error contribution to the COM PM. It is below the final combined uncertainty from random and other systematic terms (especially for M31, because of its larger area), and therefore not explicitly included in the discussion that follows.

The large-scale component can be both characterized and partially corrected using the measured DR2 PMs of quasars. \citetalias{lindegren18} find that this component has a local dispersion of $28 \uasyr$, and a characteristic correlation scale length of $20$ deg. To correct the measured COM PM $\mu_{\rm obs}$ of one of our target galaxies for this systematic error, we calculate the average PM $\mu_{\rm qso}(R)$ of the quasars within a radius $R$, and then evaluate $\mu_{\rm cor} \equiv \mu_{\rm obs} - \tau \mu_{\rm qso}(R)$. In PM studies with \textit{HST} \citep[e.g.][]{sohn12}, one observes target stars and background objects in the same small field, and it is appropriate to use $\tau = 1$. In the present case, we need to average the quasar PMs over an area that exceeds that of the target object, so as to reduce the random uncertainty in $\mu_{\rm qso}(R)$ to an acceptable level. At the same time, we need to keep $R$ as much as possible below the characteristic correlation scale length of $20$ deg, so that the average actually approximates the local systematic error. Based on the analytic treatment in Appendix A, we adopt $R = 10$ deg and $\tau = 0.93$, and show that this reduces the systematic PM error from $28$ to $16 \uasyr$. 

We have found the results obtained with this methodology to be robust against reasonable changes in the underlying assumptions, to within the final measurement uncertainties. This includes reasonable changes in the selection of the samples of target stars and quasars. Also, the results are insensitive to the details of our rotating disk models. If we fix $V_{\rm rot}$ a priori to values implied by LOS velocity studies, instead of treating it as a free parameter, then the results for the COM PM do not change appreciably. In fact, just taking the weighted average of the observed PMs, with no disk modeling at all, yields results that are consistent with the COM PM estimates from the disk model fits. This is because the target stars for both galaxies are distributed fairly symmetrically around the COM.

\subsection{M31 Results and Literature Comparison}

Figure~\ref{fig:PMrotation}a shows the predictions of the best-fit model for M31. The COM PM is shown in the inset on the bottom left. Line segments in blue show the model predictions  for individual stars brighter than G=18.5, after subtraction of the COM PM.

The best-fit model has $V_{\rm rot} = -206 \pm 86 \kms$. This amplitude is consistent with the rotation curve inferred from LOS velocity studies, which rises to $V_{\rm rot} \approx 250 \kms$ at the radii where most of the DR2 sources in M31 are located  \citep[e.g.][]{cor10}. The minus sign indicates that the rotation is counterclockwise as seen on the sky. This is consistent with expectations, given that: (a) LOS velocities are approaching on the South-West side of the disk; and (b) dust lane morphologies imply that the near side of the disk is on the North-West side (\citetalias{vdM12b}, Table 1, and references therein).

The best-fit model has COM PM ${\vec \mu}_{\rm obs} \equiv (\mu_{\alpha*}, \mu_{\delta}) = (60 \pm 14 , -24 \pm 12) \uasyr$. The average PM of the quasar sample is ${\vec \mu}_{\rm qso} = (-6 \pm 12 , 35 \pm 8) \uasyr$. Hence, the corrected PM of M31 is
\begin{eqnarray}
\label{PMGaiaAnd}
  {\vec \mu}_{\rm M31,DR2} = ( 65 \pm 18 [{\rm rand}]
                               \pm 16 [{\rm syst}] , \nonumber \\ 
                        -57 \pm 15 [{\rm rand}]
		               \pm 16 [{\rm syst}] ) \uasyr .
\end{eqnarray} 
This can be compared\footnote{Observed proper motions from Gaia, HST, and VLBA pertain to different tracer objects in different fields, and these should therefore not be compared directly. To enable a fair comparison, we consider only the implied COM PMs of each galaxy. These were obtained in each case (by us or previous authors) upon correcting the observed PMs using a model for the internal kinematics that is appropriate for the given tracer objects.} to the weighted-average PM measured by \textit{HST} for three fields, corrected for internal M31 kinematics as in vdM12: 
\begin{equation}
  {\vec \mu}_{\rm M31,HST} = ( 45 \pm 13 ,
                        -32 \pm 12 ) \uasyr
\end{equation} 
The probability of a 2D residual between these measurements as large as implied occurring by chance is 45\% (with random and systematic errors combined in quadrature). That is, the \textit{Gaia} DR2 and \textit{HST} measurements are statistically consistent at $0.8\sigma$ (the equivalent probability for a 1D Gaussian). Given that the measurements are consistent, one can take a weighted average to obtain the improved estimate
\begin{equation}
\label{PMGaiaHST}
  {\vec \mu}_{\rm M31,DR2+HST} = ( 49 \pm 11 ,
                            -38 \pm 11 ) \uasyr
\end{equation} 
This is closer to the \textit{HST} than the DR2 measurement, because the former has $\sim 2$ times smaller uncertainties. 

To correct for the solar reflex motion, and obtain the PM in the Galactocentric rest frame, one must subtract the PM
\begin{equation}
  {\vec \mu}_{\rm M31,rad} = ( 39 , -22) \uasyr
\end{equation}  
that corresponds to a purely radial approach for M31 towards the MW. This implies, still in the (RA,Dec) coordinate system %
\begin{equation}
  {\vec V}_{\rm M31,DR2+HST} = (38 \pm 41 , -61 \pm 39) \kms
\end{equation}
This differs from a purely radial orbit at an equivalent 1D-Gaussian confidence of $1.3\sigma$. Assuming a flat prior in the tangential Galactocentric velocity $V_{\rm tan}$, following \citet{vdMG08}, the median and 68\% confidence region are $V_{\rm tan, DR2+HST} = 57^{+35}_{-31} \kms$. For comparison, the \textit{HST}-only measurement implies $V_{\rm tan,HST} =  36^{+39}_{-26} \kms$ and the \textit{Gaia}-only measurement implies $V_{\rm tan, DR2} =  133^{+70}_{-68} \kms$. The latter differs from a purely radial orbit at an equivalent 1D-Gaussian confidence of $1.5\sigma$.

\begin{figure}[htp]
\includegraphics[scale=0.62]{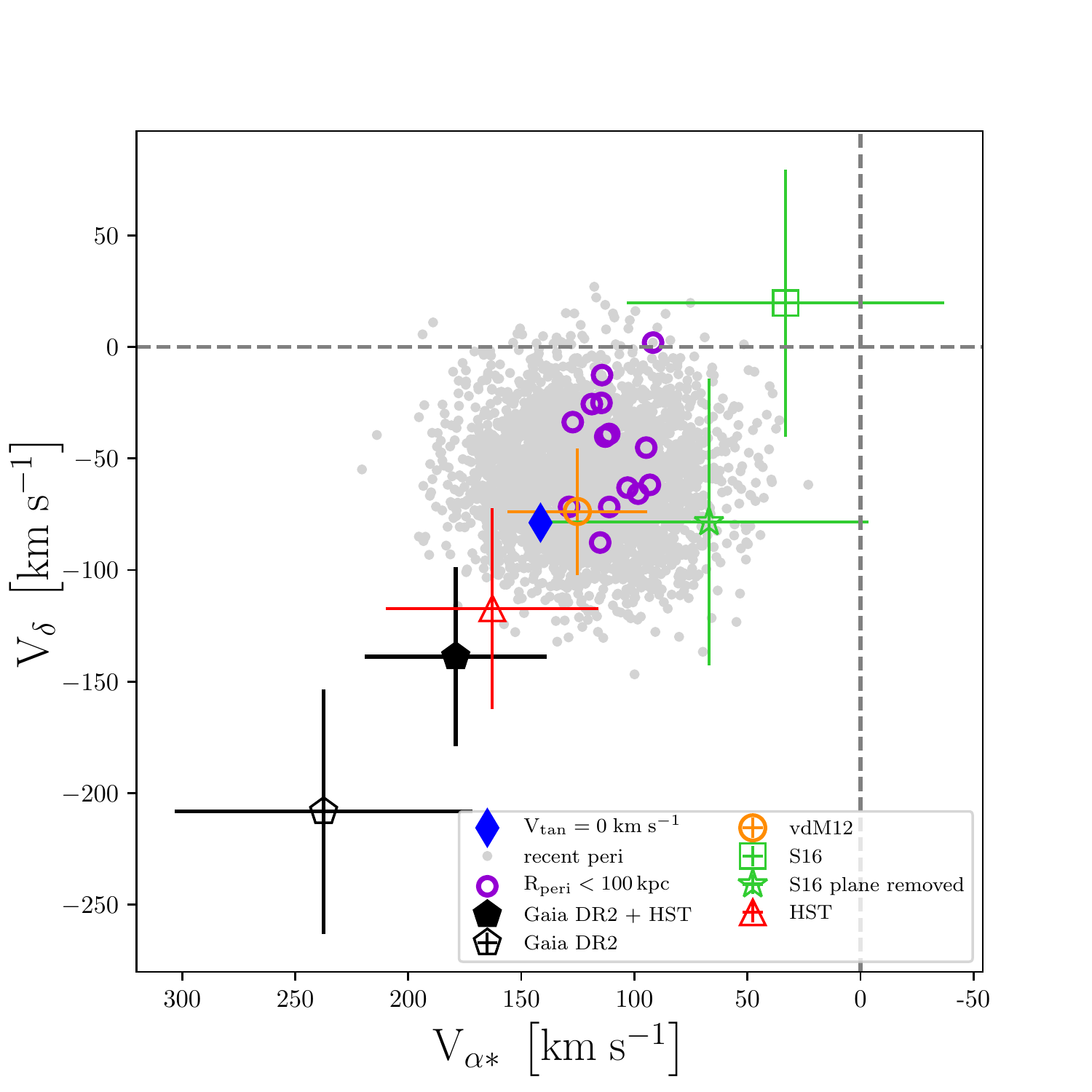}
\caption{\label{fig:M31PMs}Heliocentric M31 transverse velocity measurements $(V_{\alpha*},V_{\delta})$  (i.e., PMs in the (RA,Dec) directions transformed from $\uasyr$ to $\kms$). The blue diamond marks the transverse velocity that corresponds to a purely radial orbit for M31 towards the MW (subtraction of this velocity vector yields Galactocentric transverse velocities). Points with error bars mark the following measurements: \textit{Gaia} DR2 (open black pentagon); average \textit{HST} measurement from observations of 3 distinct fields (open red triangle; from \citetalias{vdM12a}); average of the \textit{Gaia} DR2 and \textit{HST} measurements (closed black pentagon); indirect dynamical estimates from LOS velocities of M31 satellite galaxies, with (open green square) or without (open green star) the members of the M31 plane of satellites (both from \citetalias{salomon16}); average of \textit{HST} and other indirect dynamical estimates (open orange circle; \citetalias{vdM12a}). \citetalias{patel17a} numerically calculated M33 orbits relative to M31 for velocities inside the $4\sigma$ uncertainty region for the latter average, as described in the text. The gray points indicate orbits where M33 had a pericentric approach to M31 (smaller than their current separation) in the past 6 Gyr (the ARP sample from \citetalias{patel17a}). The purple circles indicate a further subset, where the distance at pericenter was $<$ 100 kpc and the latter occurred within the last 3 Gyr (the RP100T sample from \citetalias{patel17a}). The \textit{Gaia} DR2 PM exclusively supports orbits where M33 is on first infall into M31.}
\end{figure}

\citetalias{salomon16} used an indirect dynamical method to estimate 
\begin{equation}
  {\vec V}_{\rm M31,S16} = (-112 \pm 70 , 99 \pm 60) \kms
\end{equation}
This differs from DR2+\textit{HST} weighted average at an equivalent 1D-Gaussian confidence of $2.4\sigma$. It differs by $2.5\sigma$ from the DR2 measurement itself. Figure~\ref{fig:M31PMs} compares the various measurements in the space of heliocentric (RA,Dec) velocities (i.e., transforming $\uasyr$ to $\kms$, but {\it not} correcting for the solar reflex motion). The \textit{HST} measurement is shown with a red triangle; the DR2 and DR2+\textit{HST} results are shown as open and closed black pentagons, respectively; the \citetalias{salomon16} result is shown as a green square. 

The Galactocentric velocity ${\vec v} = (V_X, V_Y, V_Z)$ of M31 implied by the DR2 measurement is
\begin{equation}
  {\vec v}_{\rm M31,DR2} =
    (0 \pm 75 , -176 \pm 51 , -84 \pm 73) \kms .
\end{equation}
The velocity implied by DR2+\textit{HST} weighted average is
\begin{equation}
  {\vec v}_{\rm M31,DR2+HST} =
    (34 \pm 36 , -123 \pm 25 , -19 \pm 37) \kms .
    \label{eq:M31weightedv}
\end{equation}
These velocities are expressed in the Galactocentric $(X,Y,Z)$ coordinate system defined in \citetalias{vdM12a}, and make the same assumptions about the solar position and velocity. 

\subsection{M33 Results and Literature Comparison}

Figure~\ref{fig:PMrotation}b is similar to Figure~\ref{fig:PMrotation}a, but now for the case of M33. The best-fit model has $V_{\rm rot} = +80 \pm 52 \kms$. This amplitude is also consistent with the rotation curve inferred from LOS velocity studies, which rises to $V_{\rm rot} \approx 100 \kms$ over the region where the DR2 sources in M33 are located \citep[e.g.][]{cor00}. The plus sign indicates that the rotation is clockwise as seen on the sky. This is consistent with expectation, given that: (a) LOS velocities are approaching on the North side of the disk; and (b) dust lane morphologies imply that the near side of the disk is on the West side (\citetalias{vdM12b}, Table1, and references therein). It is also consistent with the rotation sense inferred by \citet{brunthaler05} from the PMs of two water maser regions in M33.

The best-fit model has COM PM ${\vec \mu}_{\rm obs} = (73 \pm 14 , 32 \pm 12) \uasyr$. The average PM of the quasar sample is ${\vec \mu}_{\rm qso} = (45 \pm 13 , 66 \pm 11) \uasyr$. Hence, the corrected PM of M33 is
\begin{eqnarray}
\label{PMGaiaTri}
  {\vec \mu}_{\rm M33,DR2} = ( 31 \pm 19 [{\rm rand}]
                               \pm 16 [{\rm syst}] , \nonumber \\ 
                              -29 \pm 16 [{\rm rand}]
		               \pm 16 [{\rm syst}] ) \uasyr .
\end{eqnarray} 
This can be compared to VLBA measurements from water masers, corrected for internal M33 kinematics, in \citet{brunthaler05}
\begin{equation}
  {\vec \mu}_{\rm M33,VLBA} = ( 23 \pm 7 ,
                                8 \pm 9 ) \uasyr
\end{equation} 
The \textit{Gaia} DR2 and VLBA measurements are statistically consistent at an equivalent 1D-Gaussian confidence of $1.0\sigma$. Given that the measurements are consistent, one can take a weighted average to obtain the improved estimate
\begin{equation}
\label{PMGaiaVLBA}
  {\vec \mu}_{\rm M33,DR2+VLBA} = ( 24 \pm 7 ,
                                    3 \pm 8 ) \uasyr
\end{equation} 
This differs very little from the VLBA measurement, because that has $\sim 3$ times smaller uncertainties than the DR2 measurement. 

The Galactocentric velocity implied by the DR2 measurement alone is
\begin{equation}
  {\vec v}_{\rm M33,DR2} =
    (49 \pm 74 , 14 \pm 70 , 28 \pm 73) \kms .
\end{equation}
The velocity implied by the DR2+VLBA weighted average is
\begin{equation}
  {\vec v}_{\rm M33,DR2+VLBA} =
    (45 \pm 20 , 91 \pm 22 , 124 \pm 26) \kms .
    \label{eq:M33weightedv}
\end{equation}

If we use the weighted average values of DR2 with VLBA and \textit{HST} respectively, then the velocity vector of M33 relative to M31 has a radial component that centers around $V_{\rm rad, DR2+VLBA+HST} = -225 \kms$ and a tangential component around $V_{\rm tan, DR2+VLBA+HST} = 126 \kms$.  If instead we use only the new DR2 measurements, then $V_{\rm rad, DR2} = -209 \kms$ and $V_{\rm tan, DR2} = 85\kms$.

\section{Discussion}
\label{sec:disc}

\subsection{The Orbit of M33}

\citetalias{patel17a} performed orbital calculations for M33. Their models spanned M33 halo masses between $5-25 \times 10^{10}$ M$_{\sun}$. Two values for M31's virial mass (high mass: $2 \times 10^{12}$ M$_{\sun}$ and low mass: $1.5 \times 10^{12}$ M$_{\sun}$) were considered. M33 was modeled as an extended body and a three-component potential was adopted for M31. The present-day velocities were chosen in accordance with the \citetalias{vdM12a} PM value for M31 and with the VLBA PM value for M33, and their respective uncertainty ranges. The \citetalias{vdM12a} PM value is a weighted average of the PM measured with \textit{HST}, and several indirect dynamical estimates based on satellite LOS velocities. It is shown in Figure~\ref{fig:M31PMs} as an open orange  circle with error bars (roughly midway between the \textit{HST} PM measurement and the \citetalias{salomon16} indirect dynamical estimates). \citetalias{patel17a} calculated M33 orbits within the $4\sigma$ error ellipse for this velocity, and found that the two most likely orbital solutions are: 1) M33 is on first infall (low mass M31 model); or 2) M33 completed a long-period orbit where it made a pericentric approach around $\sim 6$ Gyr ago at a distance of $\sim$100 kpc from a high mass M31. Gray points in Figure~\ref{fig:M31PMs} show orbits that allow for a more recent ($<$ 6 Gyr ago) pericentric passage, while open purple circles show those that additionally reach within 100 kpc in the last 3 Gyr. The latter sample generally has a high mass M33 ($2.5 \times 10^{11}$ M$_{\sun}$) and M31 ($2 \times 10^{12}$ M$_{\sun}$), a mass combination that increases the odds of retrieving such an orbital solution. Both \citetalias{patel17a} and \citet{semczuk18} show that the mean \citetalias{salomon16} velocity vector does allow for a recent pericentric passage of M33 around M31, but only at distances $>$ 100 kpc. 

Using the new DR2+\textit{HST} weighted average velocity for M31 (Eq.~\ref{eq:M31weightedv}) and the DR2+VLBA weighted average velocity for M33 (Eq. \ref{eq:M33weightedv}), we numerically integrated the orbit of M33 around M31 backwards in time, following the same methodology of \citetalias{patel17a}. For the six M31-M33 mass combinations explored in \citetalias{patel17a}, the new velocities unanimously prefer a first infall orbit for M33; a long-period orbit is no longer a plausible orbital solution. The reason for this is evident from Figure~\ref{fig:M31PMs}, since the \textit{Gaia} DR2 results move the M31 PM further away from the gray points (and open purple circles) that designate a previous and recent pericenter passage. Such a first infall scenario is further supported by the study by \citet{shaya13} of the formation of planes of satellites in the Local Group, which concluded that M33's closest approach to M31 is happening now, also ruling out a possible recent tidal interaction.

\subsection{The Future Fate of the Local Group}

We next assess the impact of the new measurements on the future fate of the four most massive members of the LG: M31, the MW, M33 and LMC. We first follow the methodology outlined in \citetalias{patel17a} to model and integrate the orbits of the MW and M31 into the future, ignoring their massive satellites. We used the average DR2+\textit{HST} 
PM, and adopt two different mass ratios for the encounter: a high mass ratio encounter (M$_{\rm vir, MW} = 10^{12}$ M$_\odot$ and M$_{\rm vir, M31} = 2 \times 10^{12}$ M$_\odot$) and an equal mass ratio encounter (M$_{\rm vir, MW} = 1.5 \times 10^{12}$ M$_\odot$ = M$_{\rm vir, M31}$; compare \citet{wat18}). In both scenarios, the increased tangential velocity, relative to \citetalias{vdM12a}, is not sufficient to unbind the LG. The MW and M31 are still destined to merge. However, both the timing and the impact parameter of the first encounter have increased relative to \citetalias{vdM12b}, from $T_{\rm peri} = \sim$3.9 Gyr to $\sim$4.5 Gyr and $R_{\rm peri} \sim 31$ kpc to $\sim$130 kpc. The larger tangential velocity implied by the average DR2+\textit{HST} PM means that a future direct collision between the MW and M31 is less likely. 

We then included the dynamical influence of the LMC (M$_{\rm vir, LMC} = 10^{11}$ M$_\odot$) and M33 (M$_{\rm vir, M33} = 2.5 \times 10^{11}$ M$_\odot$) in the orbit calculations, using the \citet{k13} PM for the LMC and the DR2+VLBA PM for M33. This further delays the MW-M31 encounter time by $\sim$1 Gyr, but decreases the impact parameter by half ($\sim$75 kpc). All these calculations assume the mean 3D velocity vectors and static halo models. A more detailed analysis, searching the full PM error space, coupled with full N-body simulations of the 4-body encounter are needed to fully describe the future dynamics and merger of the MW-M31 system. 

\subsection{Cosmological Context}

The aforementioned results are broadly consistent with cosmological expectations. Using the Bolshoi dark matter only cosmological simulation, \citet{foreroromero13} find $V_{\rm tan} = 50 \pm 10$ km/s as the most probable relative tangential velocity for MW-M31 mass analogs (isolated pairs of halos with masses ranging from $7\times 10^{11}$ M$_\odot$ to $7 \times 10^{12}$ M$_\odot$ and negative relative radial velocities). In contrast, they found that only 8-12\% of cosmological MW-M31 analogs have $V_{\rm tan}/V_{\rm rad} < 0.32$, as was implied by the tangential velocity advocated by \citetalias{vdM12a}. Similar conclusions were reported by \citetalias{vdM12a}, \citet{garrisonkimmel14}, and \citet{carlesi16}. Therefore, the increase in M31's tangential motion to $V_{\rm tan, DR2+HST} = 57^{+35}_{-31} \kms$ better aligns the observational evidence with cosmological expectations. The increased tangential velocity is not sufficient to significantly increase the LG mass inferred from the Timing Argument \citep{gonzalez14}.

Also, the implied first infall orbit for M33 is consistent with cosmological expectations. \citetalias{patel17a} showed that mass analogs of M33 residing around M31-mass halos preferentially exhibit recent infall times (i.e. $<$ 4 Gyr ago). The orbits of 22\% of cosmological analogs never complete a pericentric passage about their host. Of the remaining 78\%, 32\% are able to achieve a pericentric passage at distances $<$ 55 kpc in the last 3 Gyr and the remaining 46\% complete pericentric passages but only at distances $>$ 55 kpc. At pericentric distances $\gtrsim$ 55 kpc, tidal forces can partially induce the tidal features observed in M33, but these are likely not strong enough to be the sole cause of the asymmetries in its stellar and gaseous disks. 

The main implication for a first infall M33 orbit is that its stellar and gaseous warps cannot be the result of tidal forces via a close encounter with M31. This also supports the assertions in \citetalias{patel17a} that M33 must have a significant satellite population of its own, similar to the LMC \citep{jet16,kallivayalil18}. \citet{patel18b} provide details on the predicted satellite population of M33. Multiple satellite encounters \citep[fly-bys, collisions, mergers, e.g.,][]{starkenburg16} could then have given rise to M33's warps. Other possibilities include long range tides due to M31 \citep[rather than invoking a strong tidal encounter as in][]{mcconnachie09} or that the features may be related to asymmetric gas accretion or inflows \citep[e.g.,][]{debattista99,lopez02}. Moreover, it has been shown that M33's floccuent spiral pattern and velocity field are reproducible in simulations through gravitational instabilities in the stars and gas alone \citep{dobbs18}, so it is conceivable that purely internal drivers may have contributed to the warp as well.  

\subsection{Structure of the M31 Satellite System}

We have found good agreement between the \textit{Gaia} DR2 and \textit{HST} PMs of M31, but both measurements disagree with indirect dynamical estimates of M31's PM using LOS velocities of satellite galaxies (\citet{vdMG08}, updated in \citetalias{vdM12a}, and \citetalias{salomon16}). This could be due to non-equilibrium in the M31 satellite system.  

A significant number of satellites of M31 are purportedly aligned in a kinematically coherent plane \citep{ibata13}. This coherent motion suggests that this system of satellites may not be in equilibrium with M31's dark matter halo. By contrast, for the Milky Way, \textit{Gaia} DR2 has confirmed that while a significant number of MW satellites are on polar orbital configurations, they may not be moving coherently \citepalias{gaiadr2b}. Also, a large number of satellites are found to be counter-rotating \citep{fritz18}. Furthermore, \textit{Gaia} DR2 PMs strongly suggest that some ultra-faint satellites have been accreted as satellites of the LMC \citep{kallivayalil18}. It is possible that such processes may have occurred in M31 as well, at different intervals in time (e.g. multiple group infall events). This may result in less pronounced satellite associations today, but nonetheless, could invalidate the assumption of dynamical equilibrium. 

The analysis presented in \citetalias{salomon16} provides direct support for this hypothesis. \citetalias{salomon16} repeated their analysis for the entire satellite system (open green square in Figure~\ref{fig:M31PMs}), using only the non-plane members (open green star). The result for the latter subsample, while statistically consistent with that for the full sample, is noticeably closer to the available M31 PM measurements. In fact, it agrees with the average DR2+\textit{HST} measurement at an equivalent 1D-Gaussian confidence of $1.0\sigma$, and with the \textit{Gaia} DR2 measurement by itself at an equivalent 1D-Gaussian confidence of $1.5\sigma$. It is possible that the (currently unknown) dynamical influences that created the M31 satellite plane (e.g. group infall, torques from large-scale structure, influence of prior massive accretion events) may have also distorted the kinematics of the current non-plane members. This could plausibly explain the residual differences, which are in fact barely statistically significant.

\section{Conclusions}

We have used the \textit{Gaia} DR2 to study the PMs of M31 and M33. We carefully selected samples of sources in the target galaxies with a minimum of contamination, and then analyzed their PMs using a simple rotating disk model. We used background quasars to limit the impact of residual systematics. The PM rotation of both galaxies is confidently detected, at values consistent with the known line-of-sight rotation curves.

The inferred COM motions are consistent at $0.8\sigma$ and $1.0\sigma$ with the (2 and 3 times higher-accuracy) measurements already available from \textit{HST} optical imaging and VLBA water maser observations, respectively. This lends confidence that all these measurements are robust. This is further supported by the finding that the \textit{Gaia} DR2 PM of the distant Milky Way dwarf galaxy Leo I, as determined by \citetalias{gaiadr2b} and \citet{simon18}, is consistent with the \textit{HST} measurement of \citet{sohn13} that used the same techniques as for M31. 

We used the new \textit{Gaia} PM measurements, combined with the existing measurements, to perform numerical orbit integrations. Doing this backward in time for  M33 with respect to M31, implies that M33 must be on its first infall. This is consistent with cosmological expectations, and is similar to what has been found for the LMC orbit with respect to the MW \citep{k13}. One corollary of such an orbit is that M33's stellar and gaseous warps and tails cannot be the result of tidal forces via a close encounter with M31. 

The new measurements imply that the M31 orbit towards the Milky Way is less radial than implied by the \textit{HST} measurement alone, $V_{\rm tan, DR2+HST} = 57^{+35}_{-31} \kms$. This too is in good agreement with cosmological expectations. This implies that the future collision with the Milky Way will happen somewhat later, and with larger pericenter, than previously inferred by \citetalias{vdM12b}. 

The Gaia DR2 and HST PM  measurements for M31 both differ from estimates inferred using indirect dynamical methods based on the LOS velocities of satellite galaxies. However, the agreement improves considerably when the satellites that reside in a planar configuration are removed from the sample. This suggests that non-equilibrium features in the satellite kinematics may be responsible for this discrepancy.  

The results highlight the potential of \textit{Gaia} for PM studies beyond the Milky Way satellite system, especially with future data releases. The random PM uncertainties, and many kinds of systematic uncertainties as well, decrease as the 1.5th power of the time-baseline. Therefore, the \textit{Gaia} PMs should be a factor 4.5 more accurate after the nominal mission, and a factor 12 more accurate after a possible extended mission. This will not only shed more light on the questions already addressed in the present paper, but it will also help address new questions. For example, the PMs of M31 dwarf satellite galaxies that are too faint for \textit{Gaia} can be measured with other telescopes such as \textit{HST} or the James Webb Space Telescope. Projects for such measurements are already underway or in planning. When combined with an accurate M31 PM determination from \textit{Gaia}, it then becomes possible to determine how the satellites move in 3D with respect to their parent galaxy.

\acknowledgments
This work has made use of data from the European Space Agency (ESA) mission {\it Gaia} (\url{https://www.cosmos.esa.int/gaia}), processed by the {\it Gaia} Data Processing and Analysis Consortium (DPAC, \url{https://www.cosmos.esa.int/web/gaia/dpac/consortium}). Funding
for the DPAC has been provided by national institutions, in particular the institutions participating in the {\it Gaia} Multilateral Agreement. This project is part of the HSTPROMO (High-resolution Space Telescope PROper MOtion) Collaboration\footnote{\url{http://www.stsci.edu/~marel/hstpromo.html}}, a set of projects aimed at improving our dynamical understanding of stars, clusters and galaxies in the nearby Universe through measurement and interpretation of proper motions from \textit{HST}, \textit{Gaia}, and other space observatories. We thank the collaboration members for the sharing of their ideas and software. We also thank Elena Pancino, Anthony Brown, Ulrich Bastian and the anonymous referee for useful suggestions and discussions, and Elena Sacchi for help with Figure~\ref{fig:stellarpop}.

\vspace{5mm}
\facility{Gaia, HST, VLBA}

\software{This research made use of Astropy \citep{astropy:2013, astropy:2018}, 
	scipy \citep{scipy},
	scikit-learn \citep{scikit-learn},
	astroquery \citep[][\url{https://doi.org/10.5281/zenodo.591669}]{astroquery}, 
    matplotlib \citep{matplotlib}, and numpy \citep{numpy}.
}

\appendix
\section{Correction of Large-Scale Systematics Using Quasar Proper Motions}
\label{app:quasarstats}

\citetalias{lindegren18} tested the systematic error in the DR2 reference frame by computing the two-point correlation function of the measured PMs with a sample of quasars, which should be at rest in the sky. We are not concerned here with small-scale variations over $\lesssim 1$ deg, which should average out over the size of our target galaxies (see Section 3.1). On larger scales, \citetalias{lindegren18} derive a fit of the form
\begin{equation}
\xi(\theta) = \sigma_0^2 \exp(-\theta/\theta_s) ,
\end{equation}
where the local dispersion is $\sigma_0 = 28 \uasyr$, and the scale length is $\theta_s = 20\degree$.
(Here the fluctuations are averaged over the RA and Dec dimensions, so this amplitude describes the fluctuation along one dimension.)  If we have no additional information, our measurements should cite a systematic error of mean zero and dispersion $\sigma_0$. But for our target galaxies in this paper we use a locally averaged set of quasars to decrease this uncertainty.

We regard the quasar PMs as a Gaussian field inhabiting a flat 2-D space, since $\theta_s$ is relatively small. To obtain the local quasar reference frame with any accuracy, we must average over some region, which corresponds to a filtering operation.
Here we only consider top hat filters with radius $\theta_f$.  We find the variance of the filtered field to be
\begin{equation}
\sigma_f^2 = 
\sigma_0^2 \int_0^\infty \theta_s^2 \left[1 + (\theta_s k)^2\right]^{-3/2} 
\left( \frac{2 J_1(\theta_f k)}{\theta_f k} \right)^2 k \, dk + \sigma_n^2 .
\label{sigmaf}
\end{equation}
This includes a noise variance term $\sigma_n$ that results from random errors in the DR2 quasar PMs. The variance of the unfiltered field is simply $\sigma_0^2$. The covariance between the
filtered field and the unfiltered field of interest is
\begin{equation}
C_{0f} = 
\sigma_0^2 \int_0^\infty \theta_s^2 \left[1 + (\theta_s k)^2\right]^{-3/2} 
\frac{2 J_1(\theta_f k)}{\theta_f k} k \, dk  .
\end{equation}
These integrals can be performed numerically.  

We can find the distribution of the unfiltered PM,
conditional on the observed value $\Delta_f$ of the filtered field, by inverting the covariance matrix describing the filtered and unfiltered values. The result is a Gaussian distribution with mean
\begin{equation}
\mu_{0|f} = \frac{C_{0f}}{\sigma_f^2} \Delta_f 
= \rho_{0f} \frac{\sigma_0}{\sigma_f}\Delta_f
\equiv \tau_{0|f} \Delta_f
\label{eqn.mean}
\end{equation}
and variance
\begin{equation}
\sigma_{0|f}^2 = \sigma_0^2 \left( 
1 - \frac{C_{0f}^2}{\sigma_0^2 \sigma_f^2} 
\right) 
= \sigma_0^2 ( 1 - \rho_{0f}^2 )
\label{eqn.dispersion}
\end{equation}
Here $\rho_{0f}$ is the correlation coefficient of the filtered and unfiltered fields.

For small sample regions, $\sigma_n$ will be very high compared to the true dispersion (i.e., the second term in eq.~[\ref{sigmaf}] will dominate) which will suppress $\rho_{0f}$. For large sample regions, $\rho_{0f}$ will drop well below 1 due to a lack of intrinsic correlation between small and large scales.  In both cases the variance approaches $\sigma_0^2$, since the quasars contribute no useful information.  However, if we can find a radius for which the averaged PM has both small noise and high intrinsic correlation, we can offset the local reference frame based on the quasars and reduce the uncertainty accordingly.  The ``translation coefficient'' $\tau_{0|f}$ in this equation tells us how to scale the circle-averaged quasar measurement.  We can then subtract the apparent motion of the quasar reference frame to get an unbiased PM measurement of the target galaxy of interest.  

For the density of quasars actually present in the AllWISE AGN/QSO catalog of \citet{sec15,sec16}, and using the \textit{Gaia} DR2 PM uncertainties, we find that the quasar PM uncertainty is $\sigma_n \approx 10 \uasyr$ at $10$ deg radius. We assume an uncertainty that scales inversely with the square root of the number of sources, and thus inversely with the filter radius.  In this case, the optimal radius appears to be about $10 \degree$.   The associated dispersion in the systematic error is $\sigma_{0|f} = 0.57 \sigma_0 = 16 \uasyr$.   The dispersion of the filtered field is $\sigma_f = 0.81 \sigma_0 = 23 \uasyr$.  The translation coefficient at this radius is $\tau_{0|f} = 0.93$; that is, one should not subtract the full value of the quasar average from the local PM, but only 0.93 of it.  It is $<\!1$ because the uncertainty in the quasar sample dilutes the correlation with the true local value.  We adopt these values in Section 3.1.

The filtered values actually obtained for the reference frame, near M33 in particular (see Section 3.3), are somewhat large compared to the expected dispersion $\sigma_0$. However, the quasar PM maps in \citet[][their figure 11]{mignard18}, while noisy, indicate that M31 and M33 do in fact inhabit the sky region with the highest apparent PMs, so perhaps the obtained values are telling us nothing more than that. The maps do also suggest a possibly non-Gaussian character of the noise field. So we acknowledge that the model presented here is only an approximation. Nevertheless, we believe that it provides a reasonable understanding of how quasar averages can be used to improve the local reference frame in \textit{Gaia} DR2. 

\section{Estimating small-scale systematic effects on the derived proper motion}
\label{app:smallscale}

\begin{figure*}[tp]
\gridline{
\fig{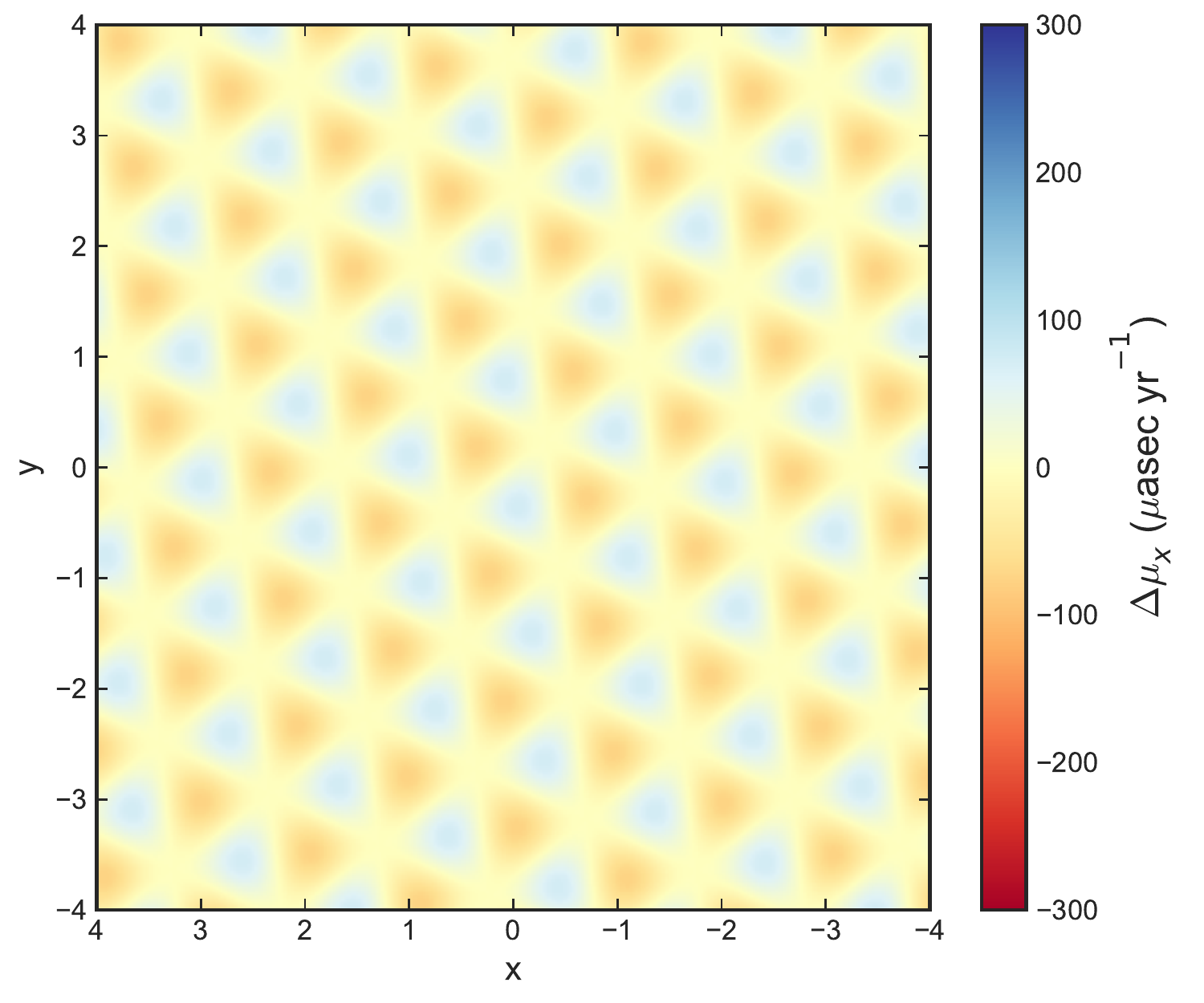}{0.333\textwidth}{(a)}
\fig{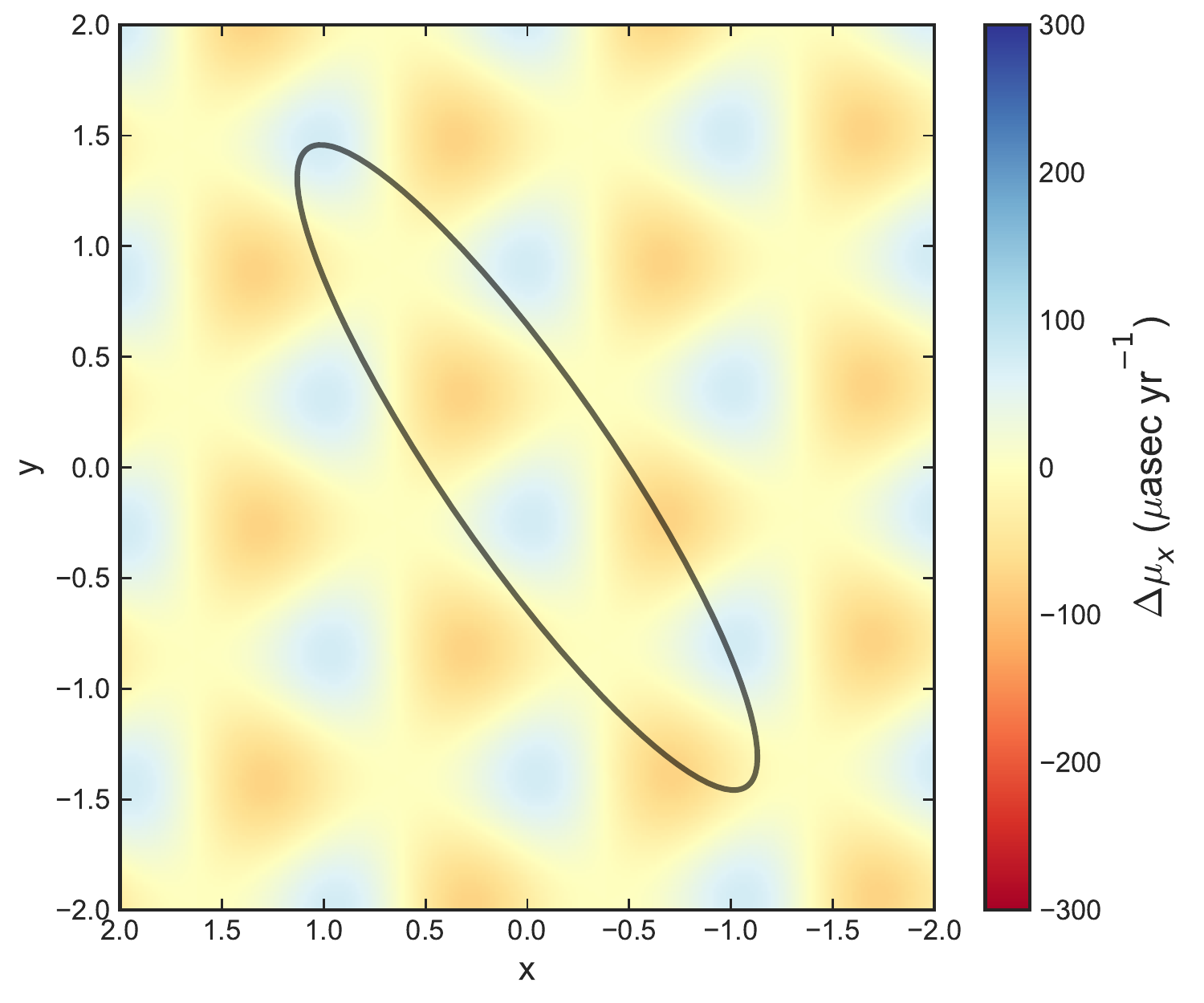}{0.333\textwidth}{(b)}
\fig{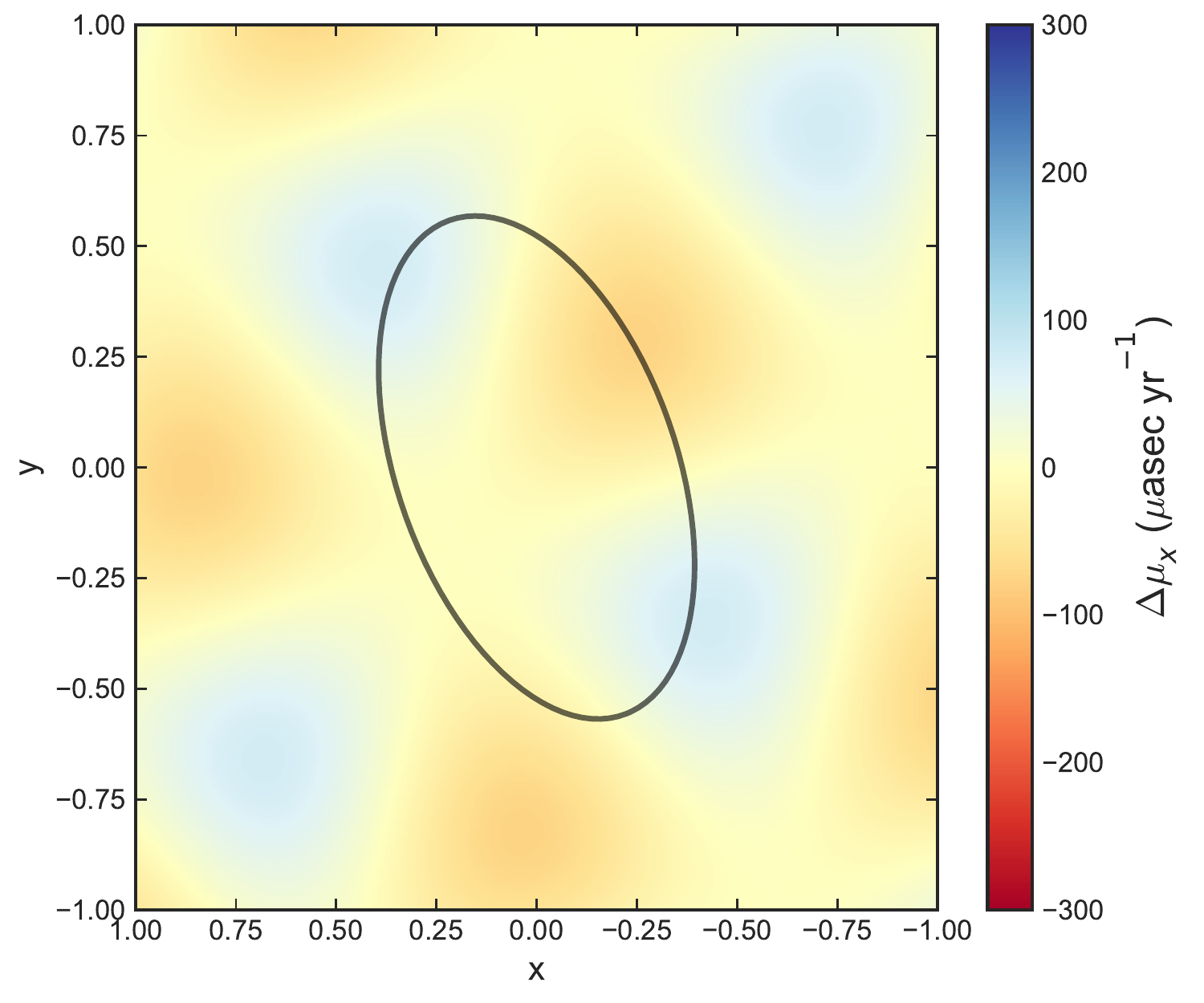}{0.333\textwidth}{(c)}}
\caption{\label{fig:smallscalenoise} 
A simple model for the small-scale pattern noise in the
\textit{Gaia} DR2 proper motion, visualized on three angular scales.
Axes represent angular offsets from the center.
(a) Uses the same angular and color scales as figure~17 of \citetalias{gaiadr2b}
(see in particular the top right panel of that figure).
(b) Plotted along with the elliptical cut we used to select M31 sources.
(c) Same but for the M33 elliptical cut.}
\end{figure*}

In the original DR2 release papers, \citetalias{lindegren18} and \citetalias{gaiadr2b} both showed that the parallax and proper motion zero-points vary in \textit{Gaia} DR2, with a regular pattern on scales of $\tsim 1 \degree$ likely related to the \textit{Gaia} scanning law (see figures~16, 17, and A10 in \citetalias{gaiadr2b}, and figure~13 in \citetalias{lindegren18}). From the autocorrelation functions shown in \citetalias{lindegren18} (figures 14 and 15), it seems reasonable to treat this as a separate effect from the large-scale errors considered in Section~\ref{sec:method} and Appendix~\ref{app:quasarstats}. \citetalias{gaiadr2b} in their section 4.1 estimate an rms amplitude of $35 \uasyr$. \citetalias{lindegren18} instead estimate an rms value of $66 \uasyr$ on small scales.  However, the latter includes the large-scale fluctuations as well, which we have explicitly modeled.  Removing the estimated $28 \uasyr$ contribution at small radius,  and using the error bar shown for the smallest point shown in their figure~15, we find a small-scale systematic amplitude of $60 \pm 13 \uasyr$, which is larger than but consistent within $2 \sigma$ with the \citetalias{gaiadr2b} estimate.

To obtain a crude estimate of the impact of these small-scale systematics on our COM PM measurements for M31 and M33, we need a model of the pattern noise. We use the product of three sine waves, each of period $2 \degree$ and rotated at angles $120 \degree$ to each other. We set the amplitude to $115 \uasyr$, which yields an rms offset of $35 \uasyr$ consistent with \citetalias{gaiadr2b}.  We then shift the origin and rotate the pattern by random amounts.  The first panel of Figure~\ref{fig:smallscalenoise} shows this pattern on a size and colorbar scale identical to that of Figure 17 in H18.  While the actual \textit{Gaia} pattern noise seems to vary depending on location and PM axis (RA or declination), our simple model seems like a reasonable approximation.  (See also Figure~13 of L18, which shows an even clearer regular pattern, although for parallax rather than proper motion.)

We then repeat our data reduction procedure numerous times, where each time we start by adding a random realization of this pattern to the proper motions of the individual sources. Figure~\ref{fig:smallscalenoise} shows examples of the pattern noise. We overplot the elliptical cuts around M31 and M33 used for sample selection in Section~\ref{sec:sample} to indicate the relevant size scale.  Clearly, the pattern noise should average out well for M31, and less well for M33 due to its smaller size. For M31 we find an rms contribution to the COM PM of only $9 \uasyr$ in each dimension, and for M33 we find $19 \uasyr$. These values are below the final combined uncertainties from random and large-scale systematic errors (eqs.~[\ref{PMGaiaAnd}] and~[\ref{PMGaiaTri}]).

These estimates should be considered crude and indicative only. Our model for the pattern noise is phenomenological, and not directly anchored to DR2 data. We have not determined the actual shift and orientation of the pattern relative to the positions of M31 and M33. Moreover, comparison of \citetalias{lindegren18} and \citetalias{gaiadr2b} shows that the \textit{Gaia} collaboration itself has not reached full consensus on the amplitude of the small-scale systematics. This indicates this is a difficult problem, and beyond the scope of the present paper. Given these considerations, and given the small expected size of the effect compared to other known error sources, we decided to not formally propagate the systematic errors from small-scale systematics into our final measurements. Doing so would not have meaningfully altered the averages 
with other measurements from HST or VLBA (eqs.~[\ref{PMGaiaHST}] and~[\ref{PMGaiaVLBA}]) that we use for the majority of our calculations in Section~\ref{sec:disc}. Future work on the \textit{Gaia} systematics may quantify the effect better and suggest further revisions of our results.

\begin{figure}[htp]
\gridline{
\fig{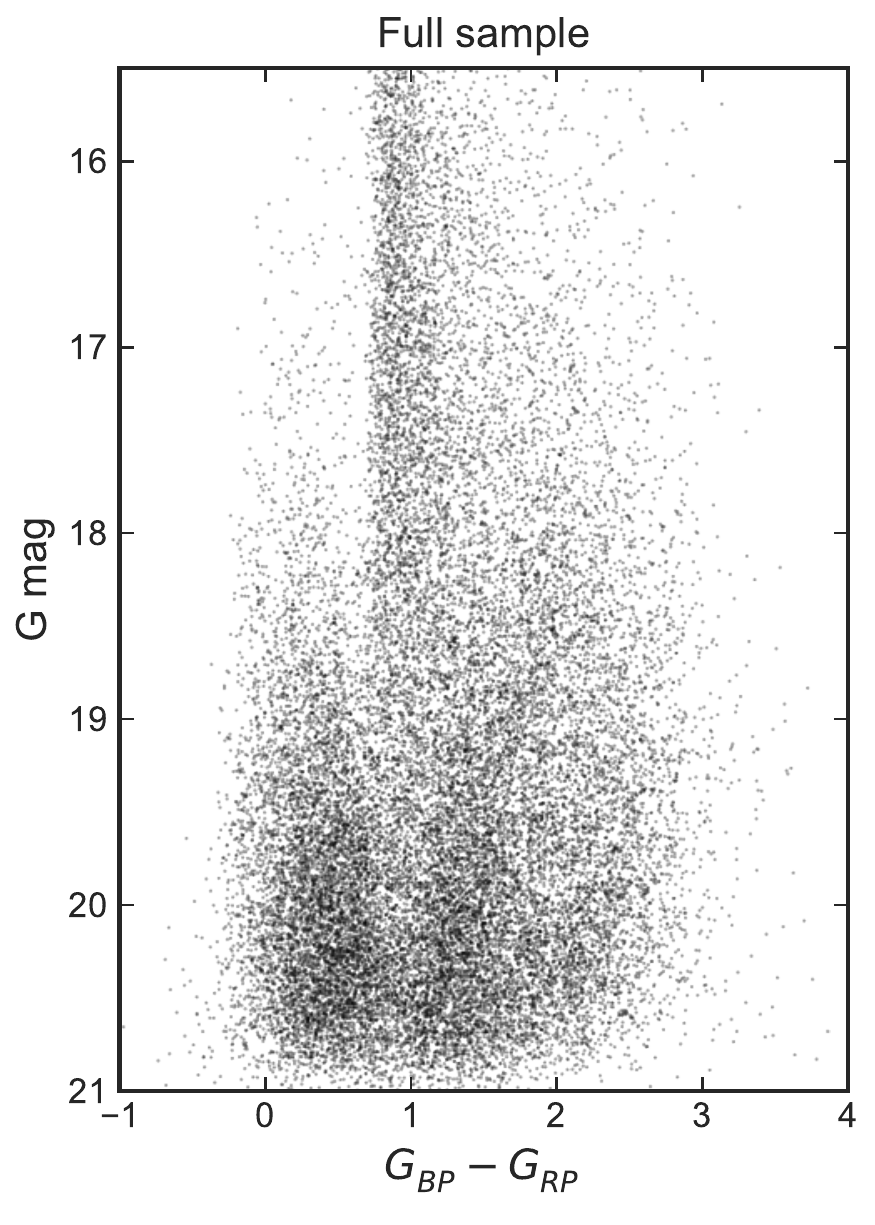}{0.25\textwidth}{(a)}
\fig{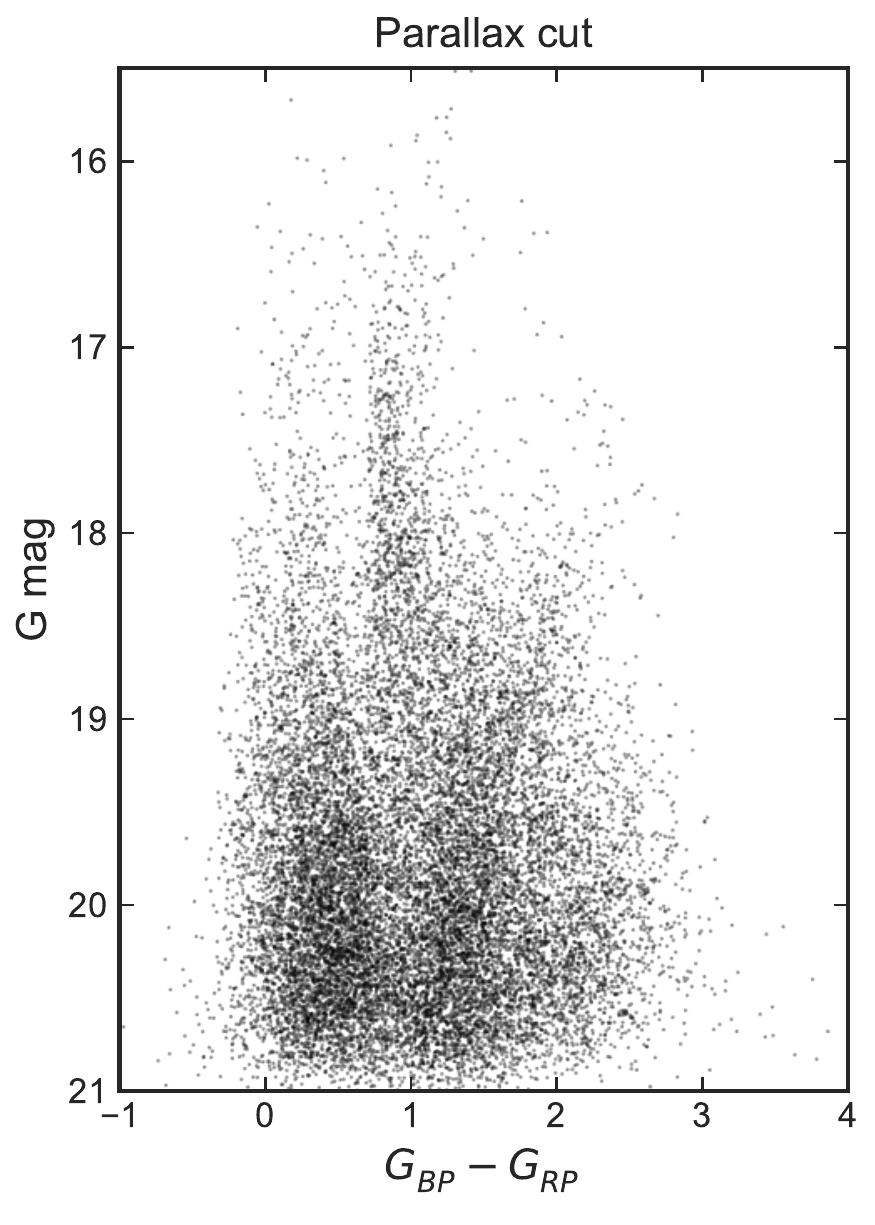}{0.25\textwidth}{(b)}
\fig{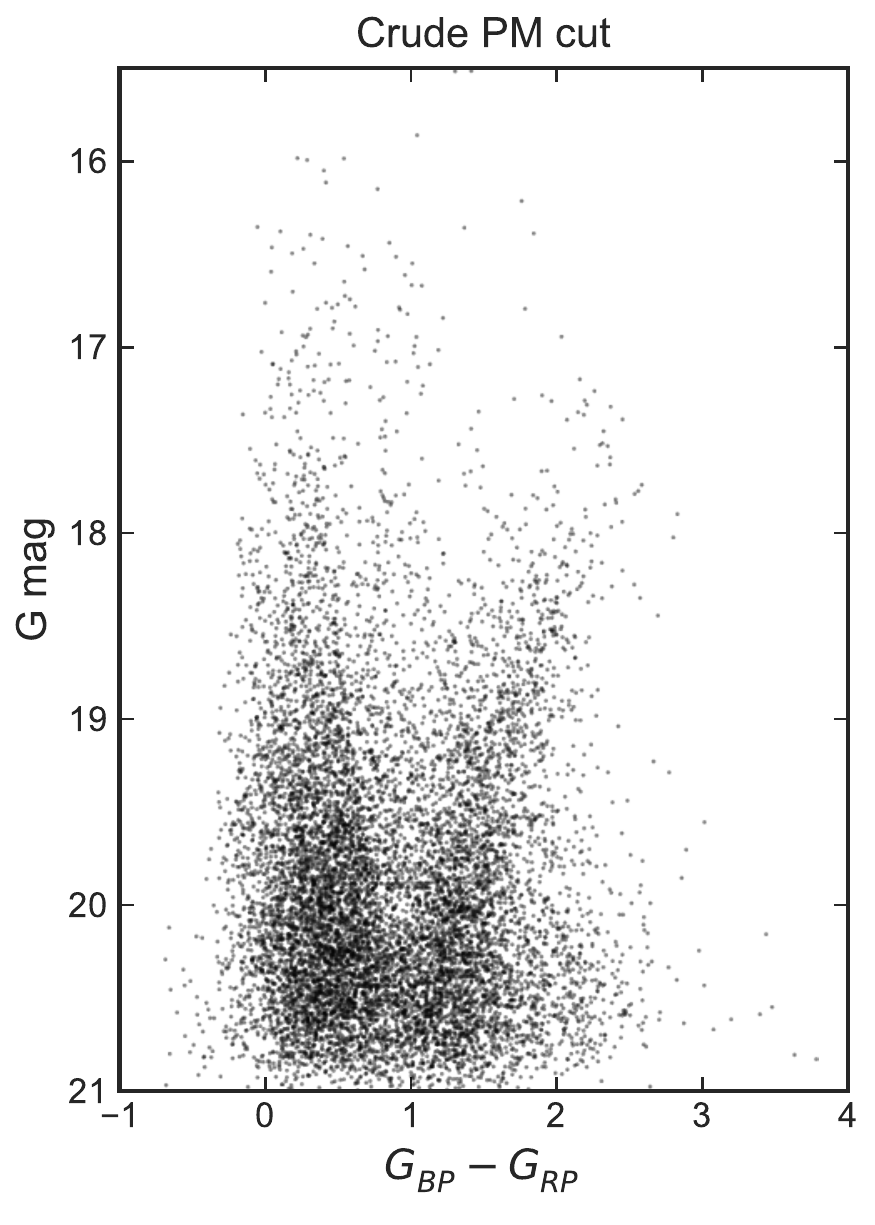}{0.25\textwidth}{(c)}
\fig{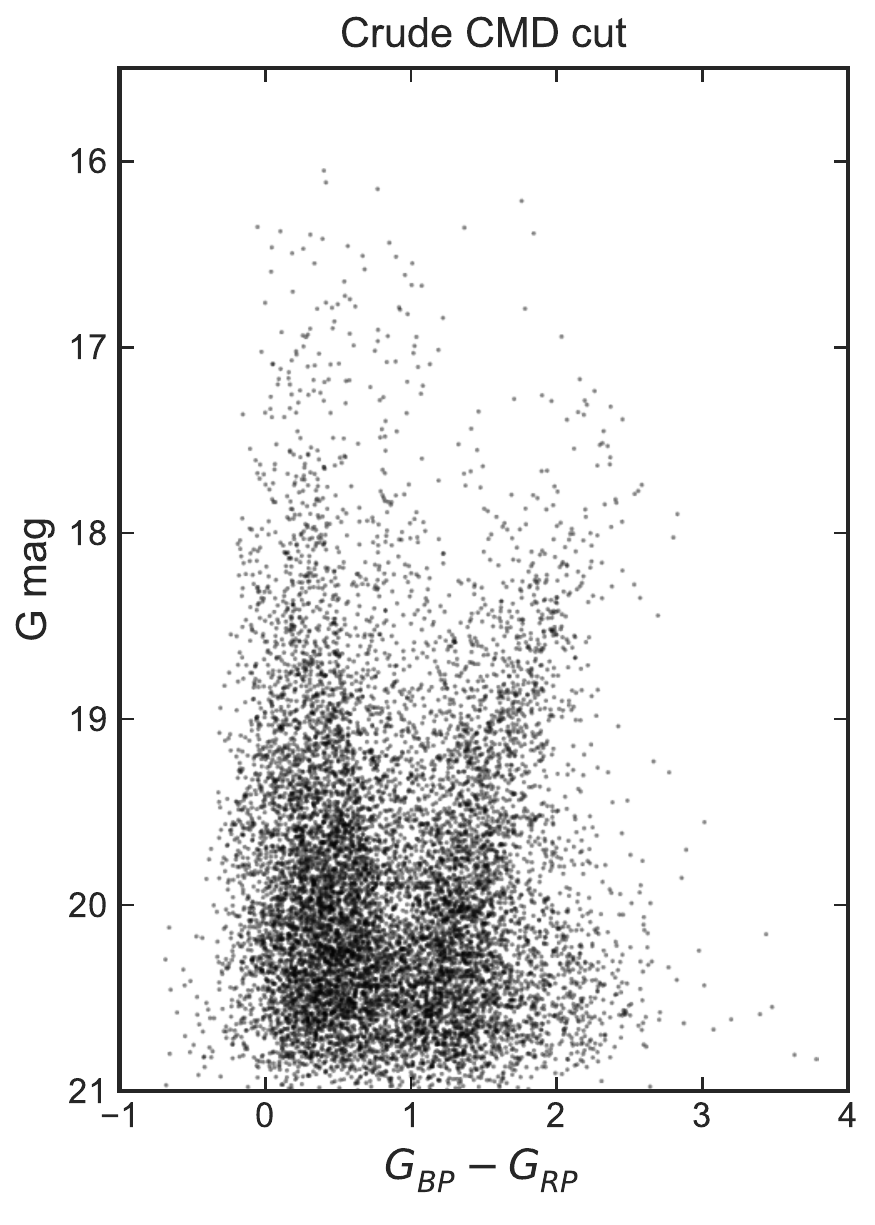}{0.25\textwidth}{(d)}}
\gridline{
\fig{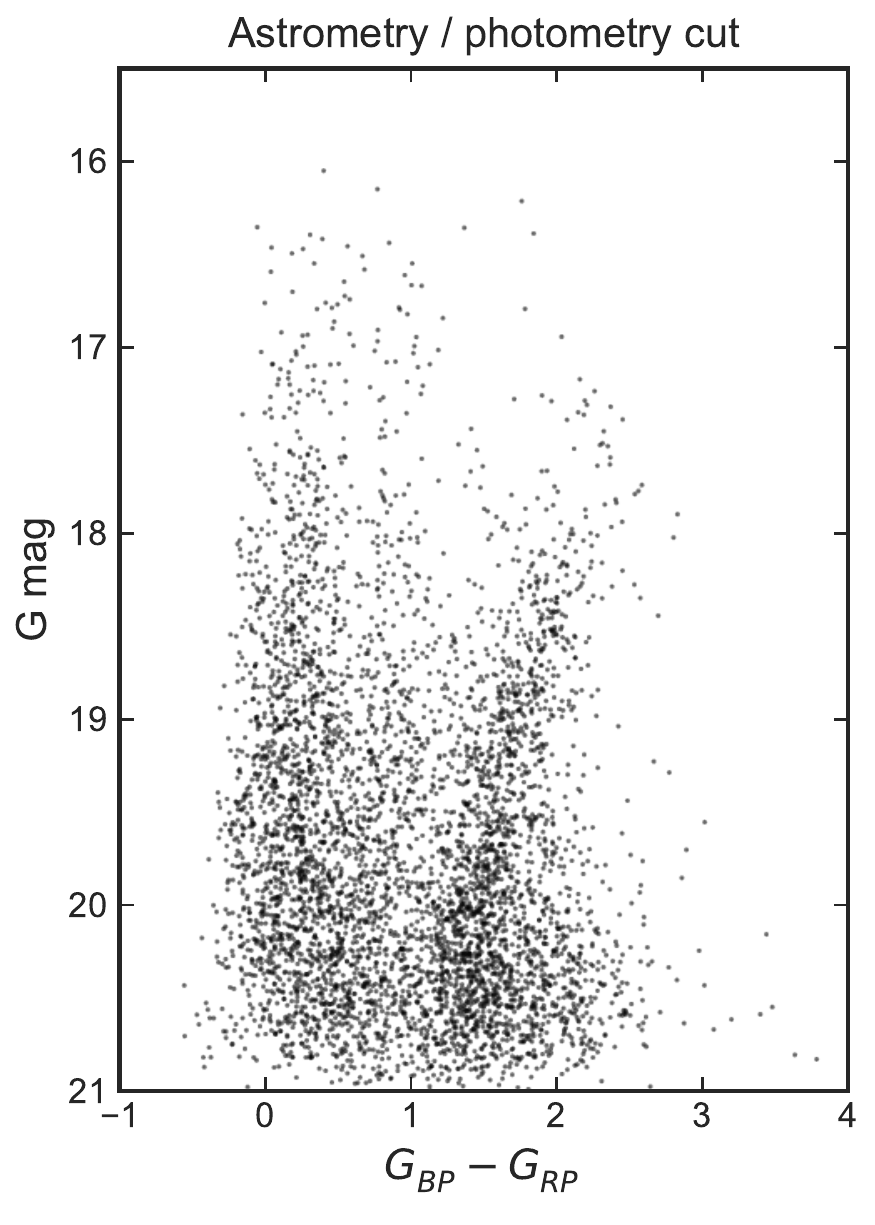}{0.25\textwidth}{(e)}
\fig{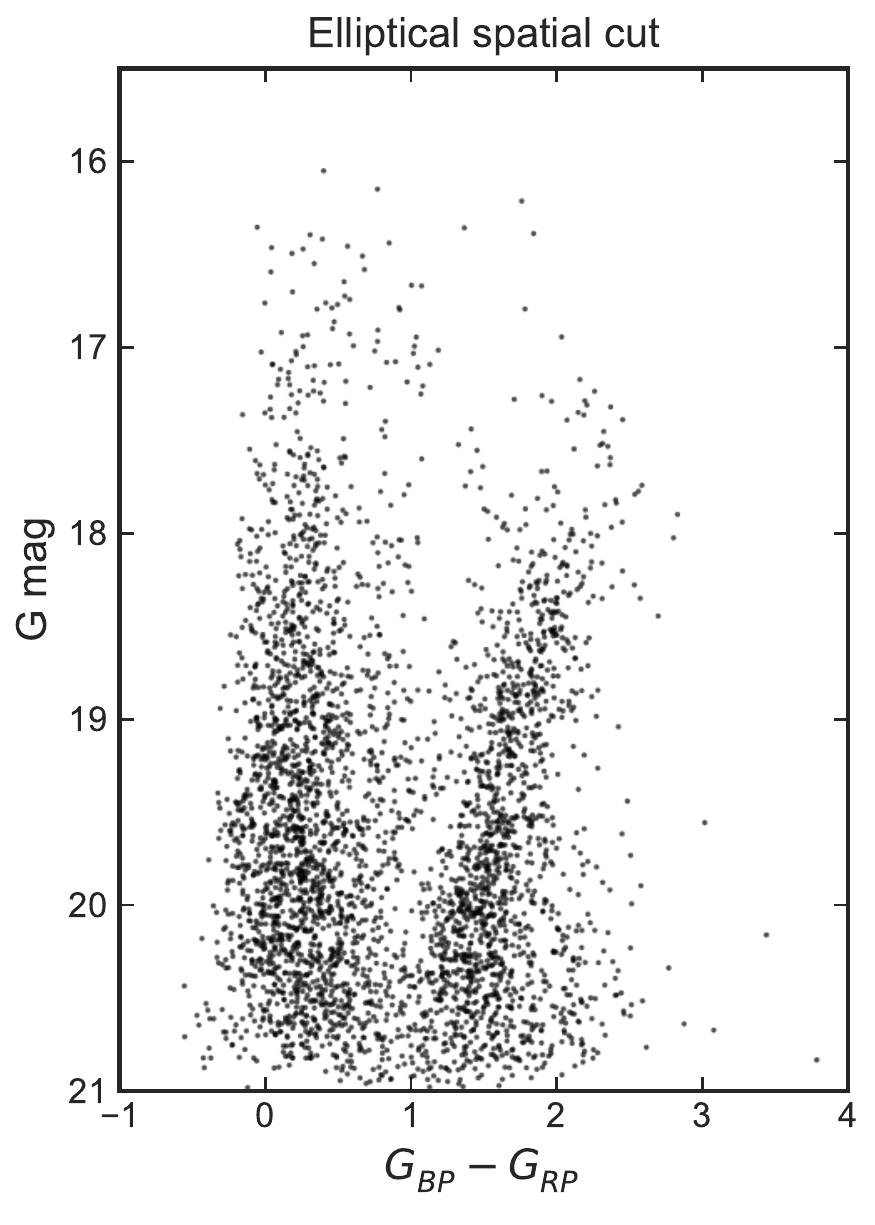}{0.25\textwidth}{(f)}
\fig{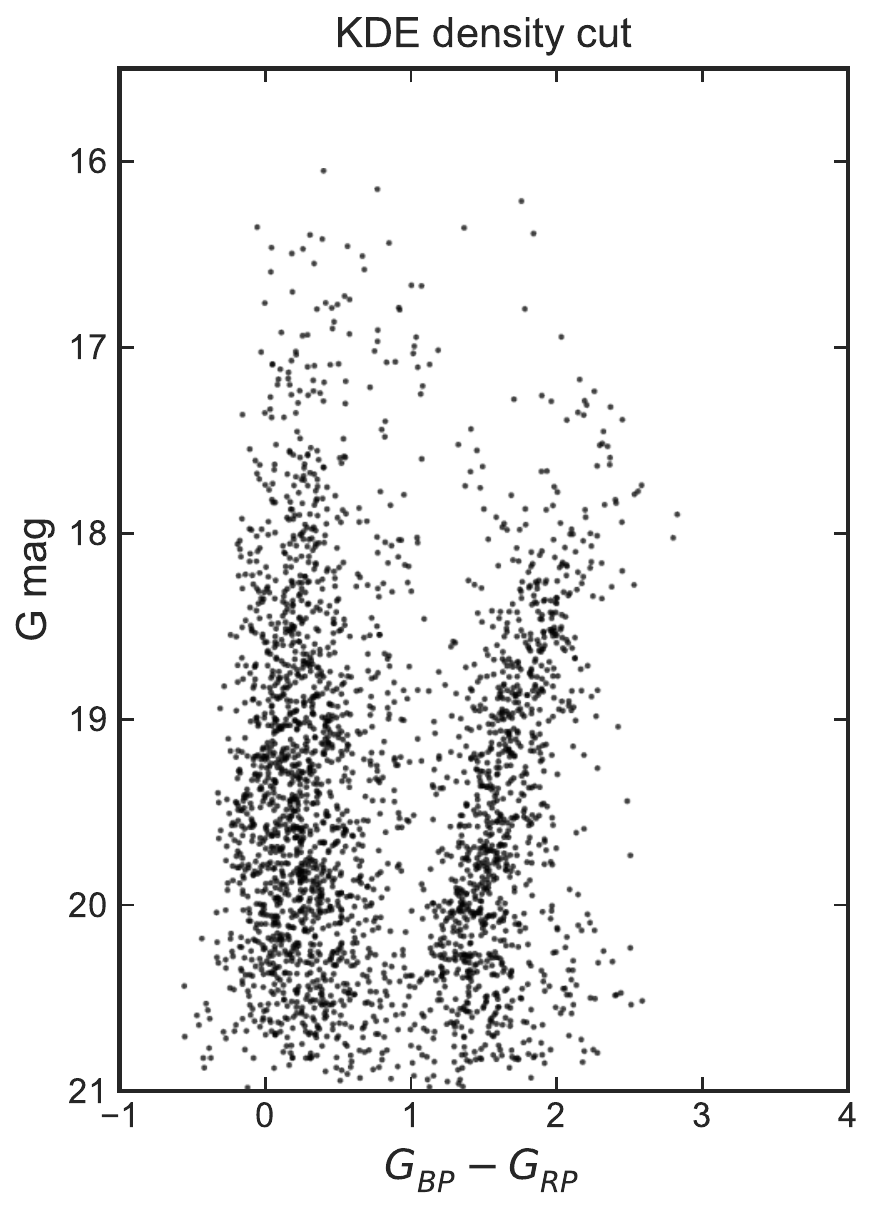}{0.25\textwidth}{(g)}
\fig{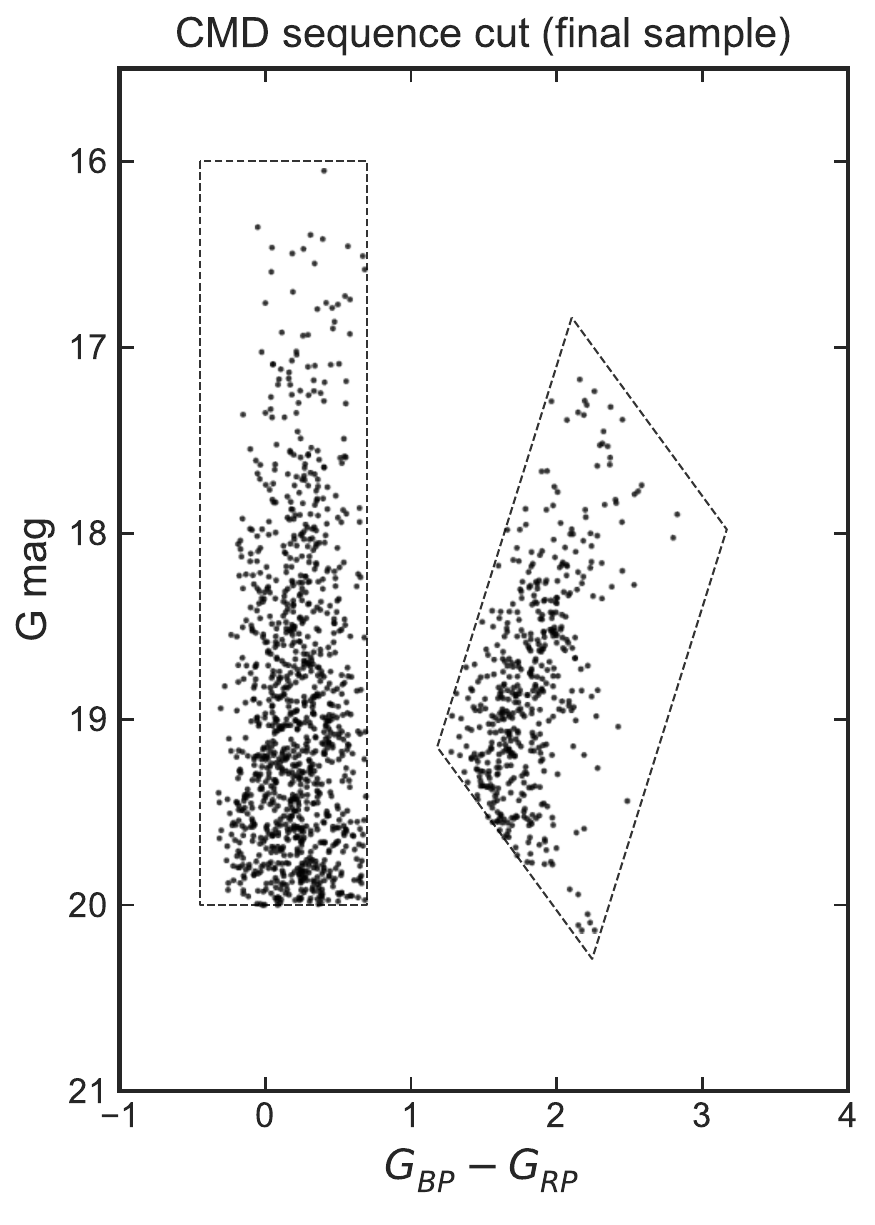}{0.25\textwidth}{(h)}}
\caption{\label{fig:cutsequence}
The panels show the effects of the various sequential source cuts described in Section~\ref{sec:sample} for the M33 sample, as follows: (a) full sample in the original circular region; (b) after applying the parallax cut; (c) after the broad proper motion cut; (d) after removing sources with $G > 16$; (e) after removing sources with bad astrometric fits and large $BP$-$RP$ excess factors; (f) after restricting to an elliptical region around the galaxy; (g) after applying the KDE spatial cut to pick out the regions most dominated by young stars in the target galaxy; and (h) final sample, after selecting within blue and red polygonal regions in the CMD (as shown) where target galaxy stars are concentrated. The point weight increases from panel to panel to keep the sample visible as the point number decreases.}
\end{figure}

\begin{figure}[htp]
\gridline{
\fig{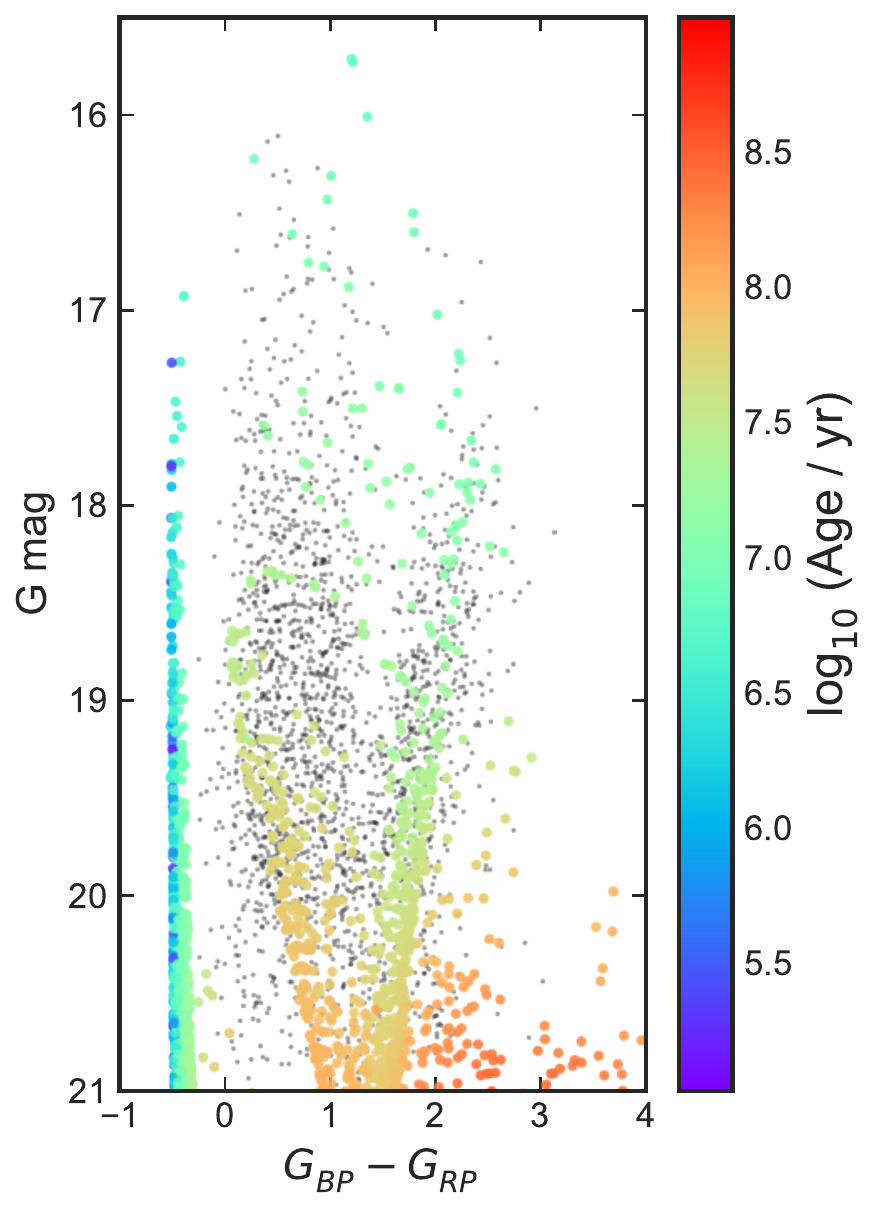}{0.25\textwidth}{(a)}
\fig{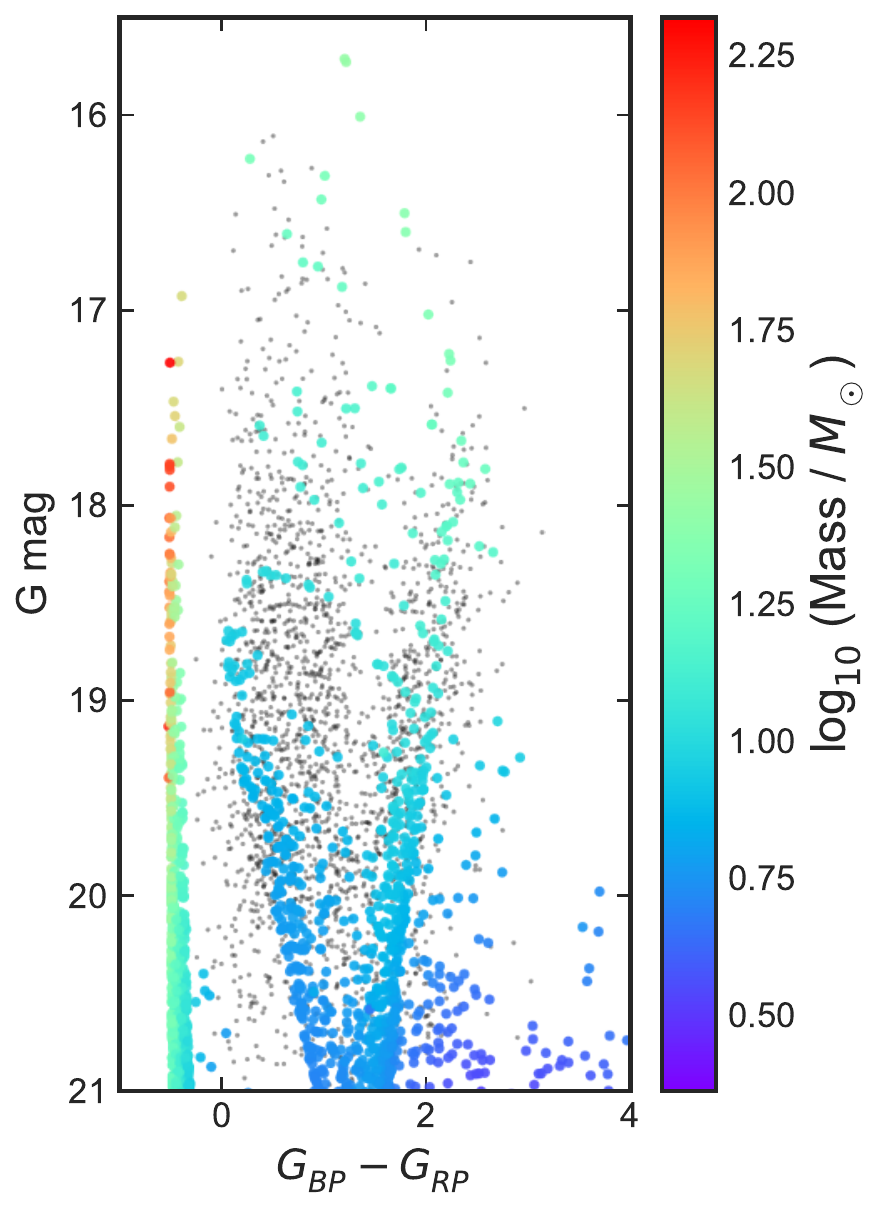}{0.25\textwidth}{(b)}
\fig{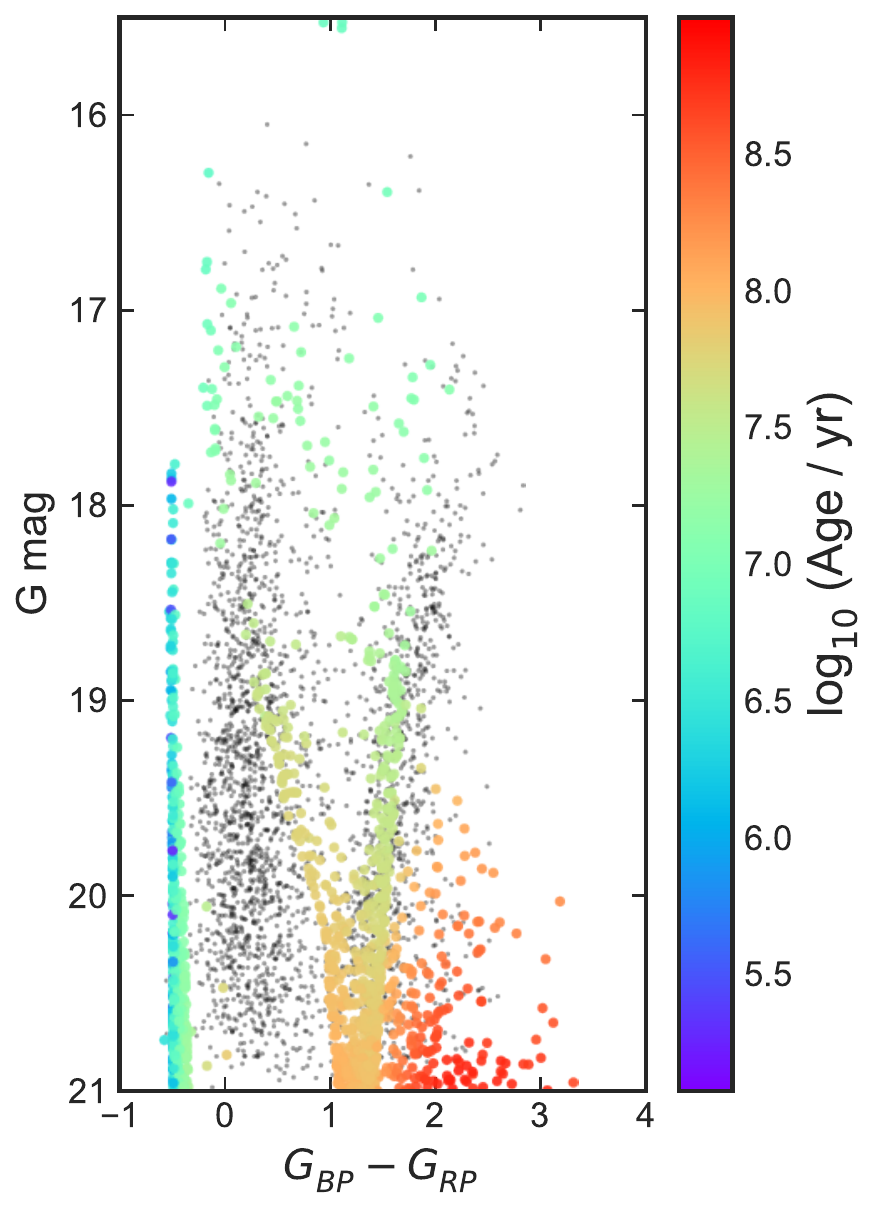}{0.25\textwidth}{(c)}
\fig{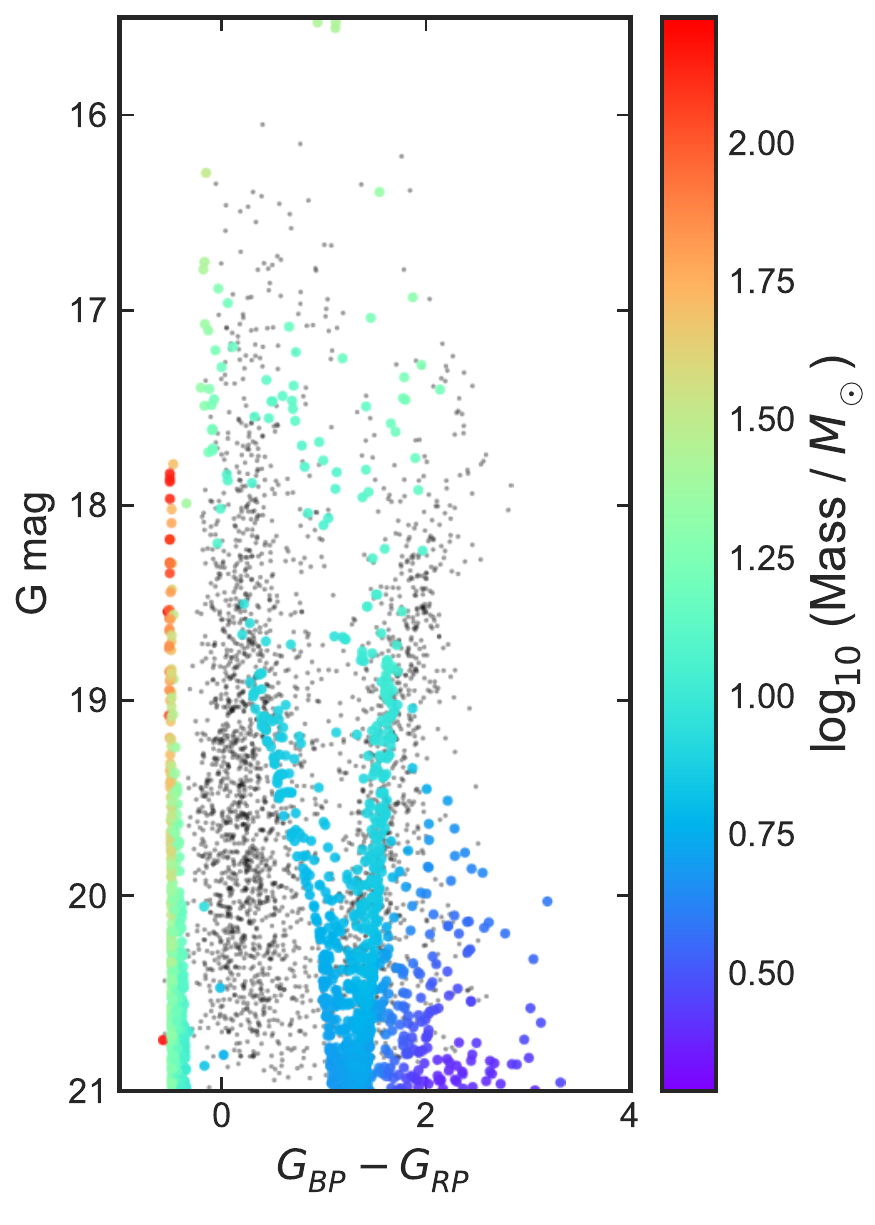}{0.25\textwidth}{(d)}}
\caption{\label{fig:stellarpop}
CMDs of artificial stellar populations, generated as described in the text, compared to the \textit{Gaia} photometric samples. Panels (a) and (b) show M31, while (c) and (d) show M33. Colored points show the artificial stars with color indicating either stellar age or stellar mass, as indicated by the colorbar. Gray points show stars selected from \textit{Gaia} DR2 using all of our sample cuts, except for the explicit CMD selection of the blue and red sequences.} 

\end{figure}

\section{Source selection and physical nature}

Figure~\ref{fig:cutsequence} illustrates how the CMD of the sample of M33 sources changes as the various selection criteria described in Section~\ref{sec:sample} are applied. The main effect is to weed out the initially strong population of foreground stars, seen as the vertical sequence extending to the brightest magnitudes shown. The M31 sample is not shown but behaves similarly.

Our chief goal in this paper is to examine the motion of stars in M31, not their intrinsic characteristics such as mass or age. Numerous studies of the Star Formation History (SFH) have already been performed for M33 \citep[e.g., ][]{davidge11} and especially M31 \citep[e.g., ][]{davidge12,lewis15,williams15,williams17}, and we do not seek to complement these here. Nevertheless, it is useful to have at least a rough qualitative understanding of the nature of the sources we are using as tracers. To this end we adopted MIST isochrones in the Gaia bands \citep{dotter16,choi16} and then used the codes described in \citet{Sacchi16} to draw artificial stellar populations. Since we are interested only in qualitative comparisons, we assumed a constant SFH between $10^5$ and $10^9 \yr$; assumed a simple Salpeter initial mass function; and ignored any effect of reddening within the host galaxies. We used metal mass fractions $Z=0.0142$ for M31 and $Z=0.00582$ for M33.

Figure~\ref{fig:stellarpop} shows the results for both targets. Gray points show (similar to Figure~\ref{fig:cutsequence}g) the \textit{Gaia} DR2 sources that remain after application of all sample cuts except for the explicit CMD cuts . In Panels (a) and (c) we color-code the artificial stars by age, and in panels (b) and (d) by stellar mass. There is clear qualitative similarity between the colors and magnitudes of the observed and artificial sources. The agreement is close enough to suggest that our measurements are based mostly on young main-sequence and (blue and red) supergiant stars of age $\tsim 3 \times 10^6$--$10^8 \yr$, with masses $\tsim 5$--$30 \Msun$. However, various discrepancies can be seen as well, especially for the bluer sources. These could be addressed by modeling in detail the reddening, SFH, age-metallicity relation, completeness, blending, and photometric errors, but all this is outside the scope of the present paper. One obvious factor is the color bias in crowded regions associated with the {\tt phot\_bp\_rp\_excess\_factor} parameter,  which significantly affects the source colors especially towards the centers of the target galaxies (see discussion in Section~\ref{sec:sample}). Extended objects such as blends of stars or H~II regions could conceivably have different astrometric uncertainties and biases than point sources, but the evidence here suggests our sources are not predominantly extended.  

\bibliographystyle{aasjournal}
\bibliography{M31M33}

\end{document}